\documentclass[11pt, twoside]{article}
\usepackage{multicol}

\usepackage{blindtext} 

\usepackage[T1]{fontenc} 
\usepackage[english]{babel} 

\usepackage{geometry}
\usepackage{graphicx}
\usepackage{booktabs} 
\usepackage{enumitem} 
\setlist[itemize]{noitemsep} 
\usepackage{graphicx}

\usepackage{abstract} 

\usepackage{titlesec} 
\titleformat{\section}[block]{\large\scshape\centering}{\thesection.}{1em}{} 
\titleformat{\subsection}[block]{\large}{\thesubsection.}{1em}{} 

\usepackage{titling} 

\usepackage{authblk}
\usepackage{floatrow}

\def\acknowledgements#1{\vskip.2in\subsubsection*{Acknowledgments}#1}

\usepackage{caption}
\usepackage{amsmath}
\usepackage{arydshln}
\usepackage[numbers]{natbib} 
\usepackage{graphics}
\usepackage{siunitx}

\usepackage{algorithm}
\usepackage[noend]{algpseudocode}
\usepackage{amssymb}

\newcommand\blfootnote[1]{%
  \begingroup
  \renewcommand\thefootnote{}\footnote{#1}%
  \addtocounter{footnote}{-1}%
  \endgroup
}

\newcommand{\beginsupplement}{%
        \setcounter{table}{0}
        \renewcommand{\thetable}{S\arabic{table}}%
        \renewcommand{\thefigure}{\arabic{figure}}%
     }

\pretitle{\begin{center}\Huge\bfseries} 
\posttitle{\end{center}} 

\title{One-shot learning and behavioral
  eligibility traces in sequential decision making
  } 
\author[1,*]{Marco P. Lehmann}
\author[2]{He A. Xu}
\author[1]{Vasiliki Liakoni}
\author[2,**]{Michael H. Herzog}
\author[1,**]{Wulfram Gerstner }
\author[3,**]{Kerstin Preuschoff}
\affil[1]{\small School of Computer and Communication Sciences and Brain-Mind-Institute, School of Life Sciences, \'{E}cole Polytechnique F\'{e}d\'{e}rale de Lausanne, CH-1015 Lausanne EPFL}
\affil[2]{Laboratory of Psychophysics, School of Life Sciences, \'{E}cole Polytechnique F\'{e}d\'{e}rale de Lausanne, CH-1015 Lausanne EPFL}
\affil[3]{Swiss Center for Affective Sciences, University of Geneva, CH-1211 Gen\`{e}ve}
\affil[*]{Corresponding author: marco.lehmann@alumni.epfl.ch}
\affil[**]{Equal contribution}

\date{} 
\renewcommand{\maketitlehookd}{%
\begin{abstract}
\noindent 
\normalsize

In many daily tasks we make multiple decisions before reaching a goal. 
In order to learn such sequences of decisions, a mechanism to link earlier actions to later reward is necessary. 
Reinforcement learning theory suggests two classes of algorithms solving this credit assignment problem: 
In classic temporal-difference learning, earlier actions receive reward information only after multiple repetitions of the task,
whereas models with eligibility traces reinforce entire sequences of actions from a single experience (one-shot).
Here we show one-shot learning of sequences.
We developed a novel paradigm to \textit{directly} observe which actions and states along a multi-step sequence are reinforced after a single reward. 
By focusing our analysis on those states for which RL with and without eligibility trace make qualitatively distinct predictions, we find direct behavioral (choice probability) and physiological (pupil dilation) signatures of reinforcement learning with eligibility trace across multiple sensory modalities. 
 
\end{abstract}
}

\begin{document}
\captionsetup[figure]{labelfont={bf},name={Fig.},labelsep=period}
\captionsetup[table]{labelfont={bf},name={Table},labelsep=period}
\maketitle
\blfootnote{to appear in eLife 8:e47463 DOI: 10.7554/eLife.47463 (2019)}  

\begin{footnotesize}

\textbf{Keywords:} eligibility trace, human learning, sequential decision making, pupillometry,  Reward Prediction Error

\acknowledgements{
This research was supported by 
Swiss National Science Foundation (no. CRSII2 147636 and no. 200020 165538), 
by the European Research Council (grant agreement no. 268 689, MultiRules),
and 
by the European Union Horizon 2020 Framework Program under grant agreement no. 720270 and no. 785907 (Human Brain Project, SGA1 and SGA2)}  

\end{footnotesize}

\section{Introduction}
In games, such as chess or backgammon,
the players have to perform a sequence of many actions before a reward  is received (win, loss).
Likewise in many sports, such as tennis, a sequence of muscle movements is performed until, for example, a successful hit is executed.
In both examples it is impossible to immediately
evaluate the goodness of a single action. Hence the question arises:
How do humans learn sequences of actions from delayed reward?

Reinforcement learning (RL) models
\cite{Sutton18} 
have been successfully used to describe reward-based learning in humans 
\citep{Pessiglione06, Glaescher10, Daw11, Niv12, ODoherty17, Tartaglia17}.
In RL, an action (e.g., moving a token or swinging the arm) leads from an old state 
(e.g., configuration of the board, or position of the body) 
to a new one. 
Here we grouped RL theories into two different classes.
The first class, containing classic  Temporal-Difference algorithms (such as TD-0 \cite{Sutton88a}),  cannot support one-shot learning of long sequences, because multiple repetitions of the task are needed before reward information arrives at states far away from the goal.   
Instead, one-shot learning requires algorithms that keep a memory of past states and actions making them eligible for later, i.e., delayed reinforcement.
Such a memory is a key feature of the second class of RL theories -- called {\em RL with eligibility trace} --, which includes
algorithms with explicit eligibility traces
\citep{Sutton88a,Watkins89,Williams92,Peng96,Singh96} and related reinforcement learning models
\citep{Sutton18,Watkins89,Mnih16,Moore93,Blundell16}.

Eligibility traces are well-established in computational models \citep{Sutton18}, and supported by synaptic plasticity experiments
\citep{Yagishita14, He15, Bittner17, Fisher17, Gerstner18}. 
However, it is unclear whether humans show one-shot learning, and a direct test of predictions that are manifestly different  between the classes of RL models with and without eligibility trace has never been performed.
Multi-step sequence learning with delayed feedback \citep{Glaescher10, Daw11, Tartaglia17, Walsh11}  
offers a way to directly compare the two, because the two classes of RL models make
{\em qualitatively} different predictions.
Our question can therefore be reformulated more precisely: Is there evidence for RL with eligibility trace in the form of one-shot learning? 
In other words, are actions and states more than one step away from the goal, reinforced after a single rewarded experience? 
And if eligibility traces play a role, how many states and actions are reinforced by a single reward? 

To answer these questions, we designed a novel sequential learning task to directly observe which actions and states of a multi-step sequence are reinforced.
We exploit that after a single reward,
models of learning without eligibility traces (our null hypothesis) and with eligibility traces (alternative hypothesis) make qualitatively distinct predictions about changes in action-selection bias and in state evaluation (Fig.~\ref{fig:F1_TaskAndHyp}).
This qualitative difference in the second episode (i.e., after a single reward) allows us to draw conclusions about the presence or absence of eligibility traces independently of specific model fitting procedures and independently of the choice of physiological correlates, be it EEG, fMRI, or pupil responses. We therefore refer to these qualitative differences as 'direct' evidence.

We measure changes in action-selection bias from behavior, and changes in state evaluation from a physiological signal, namely the pupil dilation.
Pupil responses have been previously linked to decision making, and in particular to variables that reflect changes in state value such as expected reward, reward prediction error, surprise, and risk \citep{ODoherty03, Jepma11, Otero11,Preuschoff11}.
By focusing our analysis on those states for which the two hypotheses make distinct predictions after a \textit{single} reward ('one-shot') we find direct behavioral and physiological signatures of reinforcement learning with eligibility trace. 
The observed one-shot learning sheds light on a long-standing question in human reinforcement learning \citep{Daw11, Tartaglia17, Walsh11, Bogacz07b, Walsh12, Weinberg12}.

 \section{Results}
 
 \begin{figure*}
	\centering
	\includegraphics[width=0.9\textwidth]{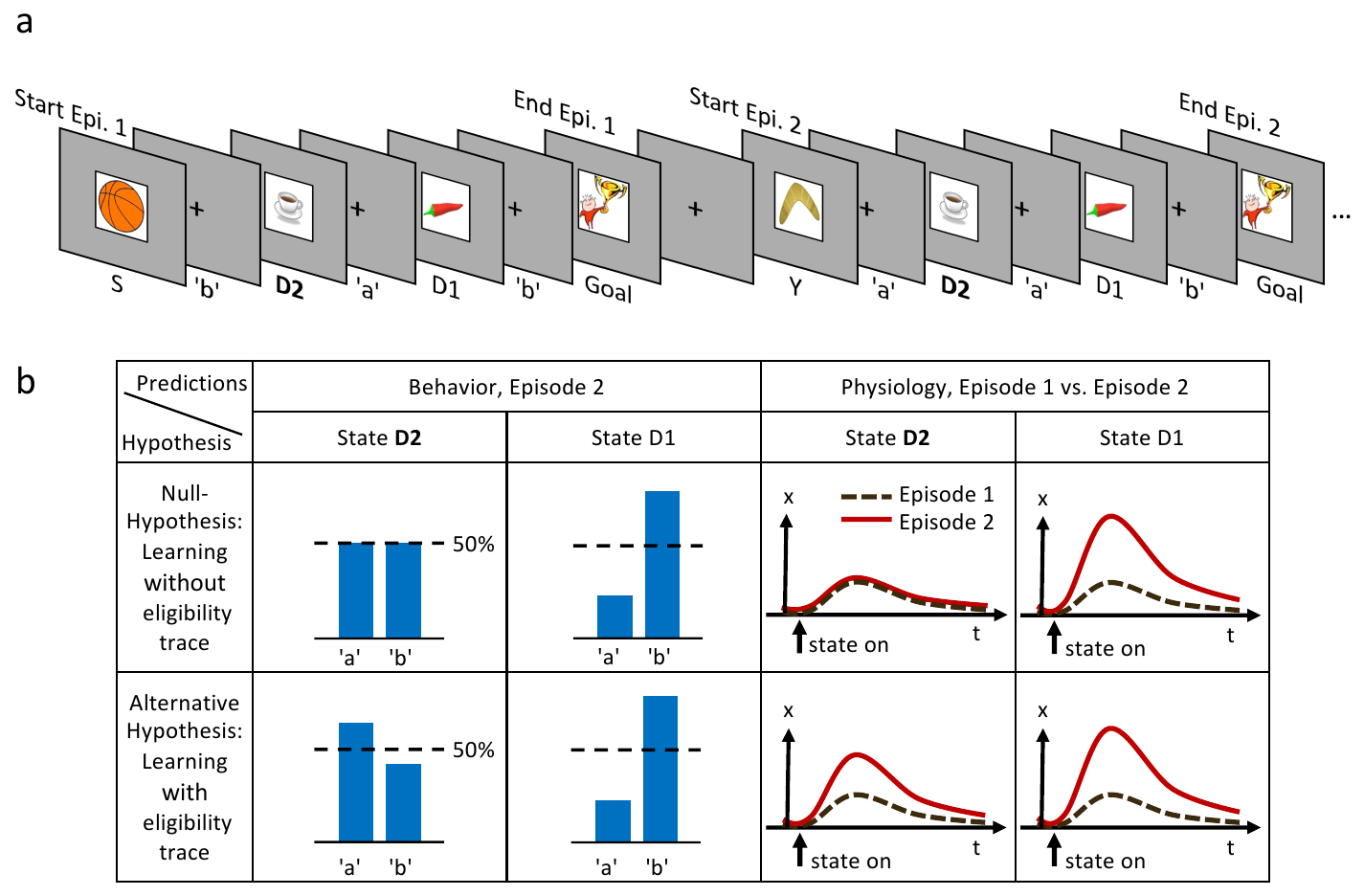}
	\caption{\textbf{Experimental design and Hypothesis:}
	\textbf{[a]} 
		Typical state-action sequences of the first two episodes. 
		At each state, participants execute one of two actions, 'a' or 'b', leading to the next state.
		Here, the participant discovered the goal state after randomly choosing three actions: 
		'b' in state S (Start), 
		'a' in D2 (two actions from the goal),
        and 'b' in D1 (one action from the goal).
        Episode 1 terminated at the rewarding goal state. 
        Episode 2 started in a new state, Y.
		Note that D2 and D1 already occurred in episode 1.
		In this example, the participant repeated the actions which led to the goal in episode 1 ('a' at D2 and 'b' at D1).
		\textbf{[b]} 
		Reinforcement learning models make  predictions about such behavioral biases,
		and about learned properties (such as action value $Q$, state value $V$ or TD-errors, denoted as $x$) presumably observable as changes in a physiological measure (e.g. pupil dilation).
		\textbf{Null Hypothesis:}
		In RL without eligibility traces, only the state-action pair immediately preceding a reward is reinforced, leading to a bias at state D1, but not at D2 (50\%-line).
		Similarly,  the state value of D2 does not change and therefore the physiological response at the D2 in episode 2 (solid red line) should not differ from  episode 1 (dashed black line).
		\textbf{Alternative Hypothesis:}
		RL with eligibility traces reinforces decisions further back in the state-action history.
		These models predict a behavioral bias at D1 and D2, and a learning-related physiological response at the onset of these states after a single reward.
		The effects may be smaller at state D2 because of decay factors in models with eligibility traces.
			}
	\label{fig:F1_TaskAndHyp}
 \end{figure*}

Since we were interested in one-shot learning, we needed an experimental multi-step action paradigm that allowed a comparison of behavioral and physiological measures between episode 1 (before any reward) and episode 2 (after a single reward).
Our learning environment had six states plus a goal G (Fig.~\ref{fig:F1_TaskAndHyp} and \ref{fig:F2_Task_Cond_Behav}),
identified by clip-art images shown on a computer screen in front of the participants.
It was designed such that participants were likely to encounter in
episode 2 the same states D1 (one step away from the goal) and/or D2 (two steps away) as in episode 1 (Fig.~\ref{fig:F1_TaskAndHyp} [a]).
In each state, participants chose one out of two actions, 'a' or 'b', and explored the environment until they discovered the goal G (the image of a reward)  which terminated the episode.
The participants were instructed to complete as many episodes as possible within a limited time of 12 minutes (Methods).

The first set of predictions applied to the state D1 which served as a control if participants were able to learn, and assign value to, states or actions.
Both classes of algorithms, with or without eligibility trace,
predicted
that effects of learning after the first reward should be reflected in the action choice probability during a subsequent visit of state D1 (Fig.~\ref{fig:F1_TaskAndHyp}[b]. For simulated data see Fig.~\ref{fig:Pupil_low_quality_traces}).
Furthermore,
any physiological variable that correlates with variables of reinforcement learning theories, 
such as  action value Q, state value V, or TD-error, should increase at the second encounter of D1.
To assess this effect of learning, we measured the pupil dilation, a known physiological marker for learning-related signals \citep{ODoherty03, Jepma11, Otero11,Preuschoff11}.
The advantage of our hypothesis-driven approach was, that we did not need to make assumptions about the neurophysiological mechanisms causing pupil changes.
Comparing the pupil dilation at state D1 in episode 1 to episode 2 (Fig.~\ref{fig:F1_TaskAndHyp}[b], null hypothesis \textit{and} alternative), provided a baseline for the putative effect.

Our second set of predictions concerned state D2.
RL without eligibility trace (null hypothesis) such as TD-0, predicted that the action choice probability at D2 during episode 2 should be at 50 percent, since information about the reward at the goal state G cannot "travel" two steps. 
However, the class of RL with eligibility trace (alternative hypothesis) predicted an increase in the probability of choosing the correct action, i.e., the one leading toward the goal (see Fig.~\ref{fig:PSA_ModelPrediction} for an estimation of the effect size).
The two hypotheses also made different predictions about the pupil response to the onset of state D2.
Under the null hypothesis, the evaluation of the state D2 could not change after a single reward.
In contrast, learning with eligibility trace predicted a change in state evaluation, presumably reflected in pupil dilation (Fig.~\ref{fig:F1_TaskAndHyp}[b]).

Participants could freely choose actions, but in order to maximize 
encounters with states D1 and D2,
we assigned actions to state transitions 'on the fly'.
In the first episode, all participants started in state $S$ (Figs. \ref{fig:F1_TaskAndHyp} and \ref{fig:F2_Task_Cond_Behav}[a])
and chose either action $a$ or $b$.
Independently of their choice and unbeknownst to the participants,
the first action brought them always to state D2, two steps away from the goal.
Similarly, in D2, participants could freely choose an action but always transitioned to D1,
and with their third action, to G.
These initial actions 
determined the assignment of state-action pairs to state transitions for all remaining episodes in this environment.
For example, if, during the first episode, a participant had chosen action $a$ in state D2 to initiate the
transition to D1, then action $a$ brought this participant in all future encounters of D2 to D1 whereas action $b$ brought her from D2 to Z (Fig.~\ref{fig:F2_Task_Cond_Behav}).
In episode 2, half of the participants started from state Y. 
Their first action always brought them
to D2, which they had already seen once during the first episode. 
The other half of the participants started in state X and their first action brought them to D1 (Fig.~\ref{fig:F2_Task_Cond_Behav}[b]). 
Participants who started episode 2 in state X started episode 3 in state Y and {\em vice versa}.
In episodes 4 to 7, the starting states were randomly chosen from $\{$S, D2, X, Y, Z$\}$.
After 7 episodes, we considered the task as solved, and the same procedure started again in a new environment (see Methods for the special cases of repeated action sequences).
This task design allowed us to study human learning in specific and controlled state sequences, without interfering with the participant's free choices.

 \subsection{Behavioral evidence for one-shot learning}
\begin{figure*}
	\centering
\includegraphics[width=0.60\textwidth]{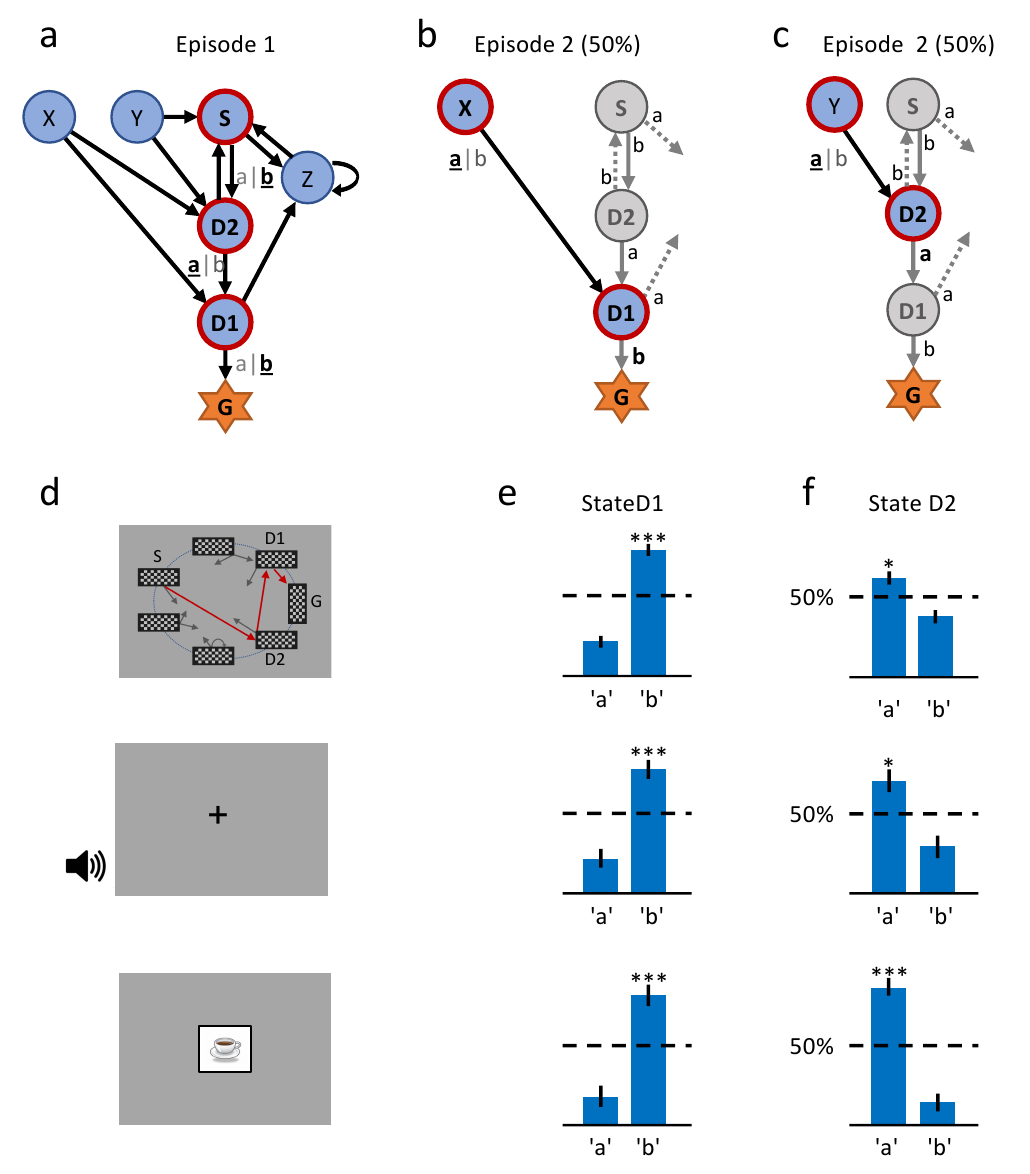}
	\caption{ \textbf{A single delayed reward reinforces state-action associations.}
	\textbf{[a]}
		Structure of the environment: 6 states, 2 actions, rewarded goal 'G'.
	Transitions (arrows) were predefined, but actions were attributed to transitions {\em during} the experiment.
	Unbeknownst to the participants, the first actions always led through the sequence  'S' (Start), 'D2' (2 steps before goal), 'D1' (1 step before goal) to 'G' (Goal).
	Here, the participant chose actions 'b', 'a', 'b' (underlined boldface).
	\textbf{[b]}	Half of the experiments, started episode 2 in X, always leading to D1, where we tested if the action rewarded in episode 1 was repeated.
	\textbf{[c]}
	In the other half of experiments, we tested the decision bias in episode 2 at D2 ('a' in this example) by starting from Y.
	\textbf{[d]}
	 The same structure was implemented in three conditions.
	 In the \textit{Spatial} condition (22 participants, \textit{top} row in Figures [d], [e] and [f]),
	 each state is identified by a fixed location (randomized across participants) of a checkerboard, flashed for a $100ms$ on the screen.
	 Participants only see one checkerboard at a time; the red arrows and state identifiers S, D2, D1, G are added to the figure to illustrate a first episode.
	 In the \textit{Sound} condition (15 participants, \textit{middle} row), states are represented by unique short sounds.
	 In the \textit{Clip-art} condition (12 participants, \textit{bottom} row), a unique image is used for each state.
	 \textbf{[e]}
	Action selection bias in state D1, in episode 2, averaged across all participants.
	 \textbf{[f]}
	In all three conditions the action choices at D2 were significantly different from chance level (dashed horizontal line) and biased toward the actions that have led to reward in episode 1.
	Error bars: SEM, $^{*} p<0.05$, $^{***} p<0.001$. 
	For clarity, actions are labeled 'a' and 'b' in [e] and [f], consistent with panels [a] - [c], even though actual choices of participants varied. 
}
	\label{fig:F2_Task_Cond_Behav}
\end{figure*}

As expected, we found that the action taken in state D1 that led to the rewarding state G was reinforced after episode 1.
Reinforcement was visible as an action bias toward the correct action when D1 was seen again in episode 2 (Fig.~\ref{fig:F2_Task_Cond_Behav}[e]).
This action bias is predicted by many different RL algorithms including the early theories of Rescorla and Wagner
\cite{Rescorla72}.

Importantly, we also found a strong action bias in state D2 in episode 2: 
participants repeated the correct action (the one leading toward the goal) in 85\% of the cases.
This strong bias is significantly different from chance level 50\% (p<0.001; Fig.~\ref{fig:F2_Task_Cond_Behav}[f]), and indicates that participants learned to assign a positive value
to the correct state-action pair after a {\em single exposure} to state D2 and a {\em single reward}
at the end of episode 1. 
In other words we found evidence for one-shot learning in a state two steps away from goal in a multi-step decision task.

This is compatible with our alternative hypothesis, i.e., the broad class of RL 'with eligibility trace',
\citep{Sutton18, Sutton88a, Watkins89,Williams92, Peng96,Singh96,Mnih16,Moore93,Blundell16}
that keep explicit or implicit memories of past state-action pairs (see Discussion).
However, it is not compatible with the null hypothesis, i.e. RL 'without eligibility trace'.
In both classes of algorithms, action biases or values that reflect the expected future reward are assigned to states.
In RL 'without eligibility trace', however,
value information collected in a single action step is shared only between neighboring states (for example between states G and D1),  whereas in RL 'with eligibility trace' value information can reach state D2 after a single episode.
Importantly, the above argument is both fundamental and qualitative in the sense that it does 
not rely on any specific choice of parameters or implementation details of an algorithm.
Our finding can be interpreted as
a signature of a behavioral eligibility trace in human multi-step decision making and complements the well-established synaptic eligibility traces observed in animal models
\citep{Yagishita14, He15, Bittner17, Fisher17, Gerstner18},

We wondered whether the observed one-shot learning in our multi-step decision task depended
on the choice of stimuli. If clip-art images helped participants to construct an imaginary story
(e.g., with the method of loci \citep{Yates66}) in order to rapidly memorize state-action associations, the effect should disappear with other stimuli.
We tested participants in environments where states were defined by acoustic stimuli (2nd experiment: 'sound' condition)
or by the spatial location of a black-and-white  rectangular grid on the grey screen
(3rd experiment: 'spatial' condition; see Fig.~\ref{fig:F2_Task_Cond_Behav} and Methods).
Across all conditions, results were qualitatively similar (Fig.~\ref{fig:F2_Task_Cond_Behav}[f]):
not only the action directly leading to the goal (i.e., the action in D1) but also
the correct action in state D2 were chosen in episode 2 with a probability significantly different from a random choice.  
This behavior is consistent with the class of RL with eligibility trace, and excludes all algorithms
in the class of RL without eligibility trace.

\begin{figure*}
	\centering
	\includegraphics[width=0.6\textwidth]{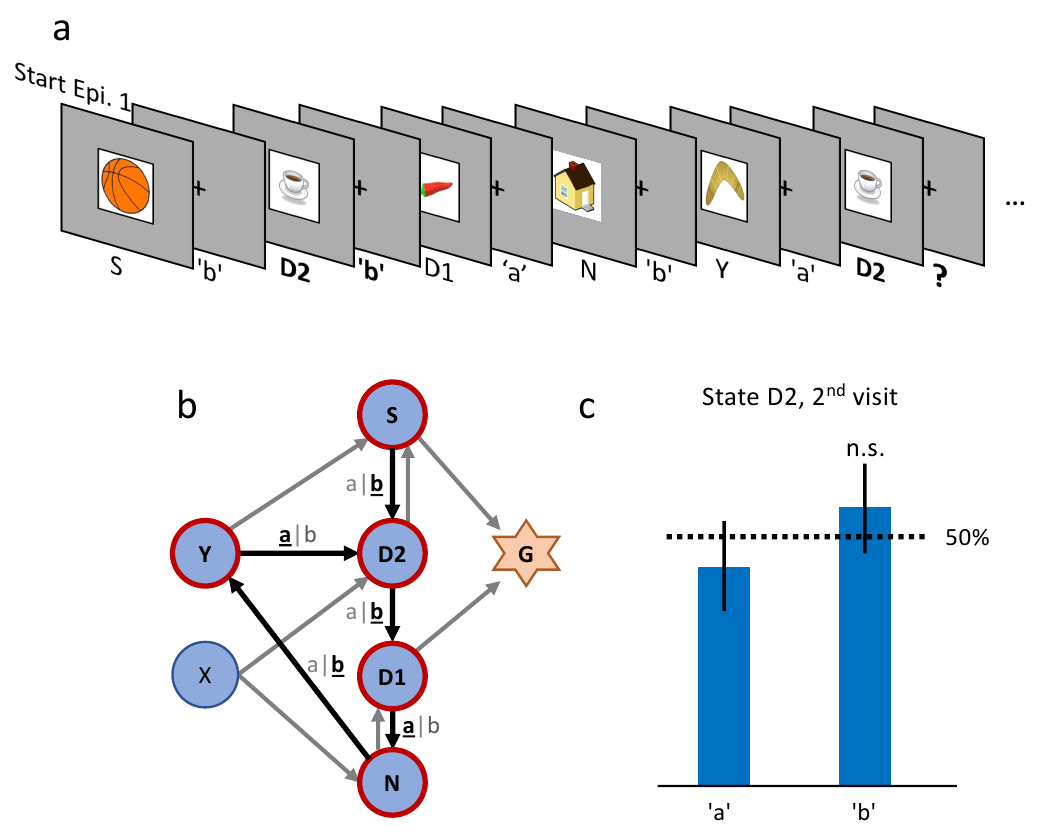}
	\caption{\textbf{Control Experiment without reward.} \\
		\textbf{[a]} Sequence of the first six state-action pairs in the first control experiment. 
		 The state D2 is visited twice and the number of states between the two visits is the same as in the main experiment. The original goal state has been replaced by a non-rewarded state N. 
		 The control experiment focuses on the behavior during the second visit of state D2, further state-action pairs are not relevant for this analysis.
		\textbf{[b]}
		  The structure of the environment has been kept as close as possible to the main experiment (Fig.~\ref{fig:F2_Task_Cond_Behav}[a]).
		\textbf{[c]} 
		Ten participants performed a total of 32 repetitions of this control experiment. 
		Participants show an average action-repetition bias of $56\%$. 
		This bias is not significantly different from the $50\%$ chance level ($p=0.45$) and much weaker than the 85\% observed in the main experiment (Fig.~\ref{fig:F2_Task_Cond_Behav}[f]).
	}
	\label{fig:Control_BehavNoReward}
\end{figure*}

Event though results are consistent across different stimuli, we cannot exclude
that participants simply memorize state-action associations independently of the rewards.
To exclude  a reward-independent memorization strategy, we performed  a control experiment in which we tested the action-bias at state D2 (see Fig.~\ref{fig:Control_BehavNoReward}) in the absence of a reward.
In a design similar to the clip-art experiment (Fig.~\ref{fig:F1_TaskAndHyp}[a]),  the participants freely chose actions that moved them through a defined, non-rewarded, sequence of states (namely S-D1-D2-N-Y-D2, see Fig.~\ref{fig:Control_BehavNoReward}[b]) during  the first episode. By design of the control experiment participants reach the state D2 twice before they encounter any reward.
  Upon their second visit of state D2, we measured whether participants repeated the same action as during their first visit.
Such a repetition bias could be explained if participants tried to memorize and repeat  state-action associations even in the absence of a reward between the two visits.  In the control experiment we observed  a weak non-significant (p = 0.45) action-repetition bias of only 56 \% (Fig.~\ref{fig:Control_BehavNoReward}[c]) in contrast to the main experiment (with a reward between the first and second encounter of state D2) where we observed a repetition bias of 85\%. These results indicate that earlier rewards influence the action choice when a state is encountered a second time.

\begin{figure*}
	\centering
	\includegraphics[width=\textwidth]{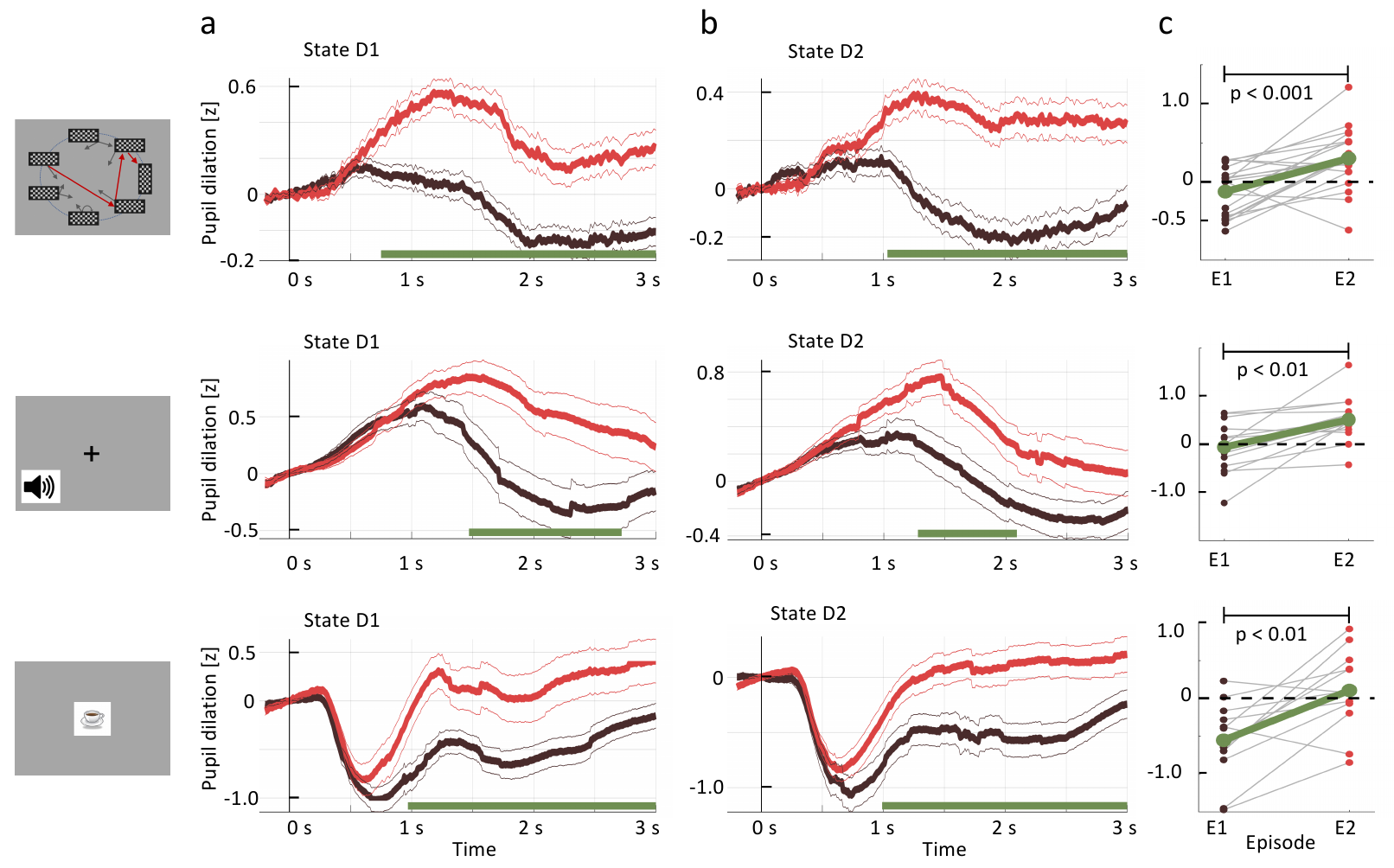}
	\caption{
	\textbf{Pupil dilation reflects one-shot learning.}
	\textbf{[a]} Pupil responses to state D1 are larger during  episode 2 (red curve)
        than during episode 1 (black). 
	\textbf{[b]} Pupil responses to state D2 are larger during  episode 2 (red curve)
        than during episode 1 (black).
	\textit{Top row}:  spatial, \textit{middle row}: sound, \textit{bottom row}: clip-art condition.
        Pupil diameter 
        averaged across all participants in units of standard deviation (z-score, see Methods),
        aligned at stimulus onset and plotted as a function of time since stimulus onset.
        Thin lines indicate the pupil signal $\pm$SEM.
	Green lines indicate the time interval during which the two curves differ significantly ($p<FDR_{\alpha}=0.05$).
        Significance was reached at a time $t_{min}$, which depends on the condition and the state:
\textit{spatial D1:} $t_{min}= 730$ ms (22, 131, 85);
\textit{spatial D2:} $t_{min}= 1030$ ms (22, 137,130)
\textit{sound D1:} $t_{min}= 1470$ ms (15, 34, 19);
\textit{sound D2:} $t_{min}= 1280$ ms (15, 35, 33);
\textit{clip-art D1:} $t_{min}= 970$ ms (12, 39, 19);
\textit{clip-art D2:} $t_{min}= 980$ ms (12, 45, 41);
(Numbers in brackets: number of participants, number of pupil traces in episode 1 or 2, respectively).
 \textbf{[c]}
	 Participant-specific mean pupil dilation at state D2 (averaged over the interval [1000ms, 2500ms]) before (black dot) and after (red dot) the first reward.
	 Grey lines connect values of the same participant. Differences between episodes are significant
	 (paired t-test, p-values indicated in the Figure).
	 }
	\label{fig:PupilDiameter_GrandAvg}
      
\end{figure*}

\subsection{Reinforcement learning with eligibility trace is reflected in pupil dilation}

We then investigated the time-series of the pupil diameter.
Both, the null and the alternative hypothesis predict a change in the evaluation of state D1, when comparing the second with the first encounter.
Therefore, if the pupil dilation indeed serves as a proxy for a learning-related state evaluation (be it Q-value, V-value, or TD-error), we  should observe a difference between the pupil response to the onset of state D1 before (episode 1) and after (episode 2) a single reward.

We extracted (Methods) the time-series of the pupil diameter, 
focused on the interval [0s, 3s] after the onset of
states D2 or D1, and 
averaged the data across participants and environments
(Fig.~\ref{fig:PupilDiameter_GrandAvg}, black traces).
We observed a significant change in the pupil dilatory response to stimulus D1 between
episode 1 (black curve) and episode 2 (red curve).
The difference was computed per time point (paired samples t-test);
significance levels were adjusted to control for false discovery rate (FDR, \cite{Benjamini95}) which is a conservative measure given the temporal correlations of the pupillometric signal.
This result suggests that participants change the evaluation of D1 after a single reward, and that this change is reflected in pupil dilation.

Importantly, the pupil dilatory response to the state D2 was also significantly stronger in episode 2 than in episode 1.
Therefore, if pupil diameter is correlated with the state value $V$, the action value $Q$, the TD-error, or a combination thereof, then the class of
\textit{RL without eligibily trace} must be excluded as an explanation of the pupil response (i.e. we can reject the null hypothesis in Fig.~\ref{fig:F1_TaskAndHyp}).

However, before drawing such a conclusion we controlled for correlations of pupil response with other parameters of the experiment.
First, for visual stimuli, pupil responses changed with stimulus luminance.
The rapid initial contraction of the pupil observed in the clip-art condition (bottom row in Fig.~\ref{fig:PupilDiameter_GrandAvg}) was a response to the 300 ms display of the images. 
In the spatial condition, this initial transient was absent, but the difference in state D2 between episode 1 and episode 2 were equally significant. 
For the {\em sound} condition, in which stimuli were longer on average (Methods), the significant separation of the curves occurred slightly later than in the other two conditions. 
A paired t-test of differences showed that, across all three conditions,
pupil dilation changes significantly between episodes 1 and 2 (Fig.~\ref{fig:PupilDiameter_GrandAvg}[c];
paired t-test, p<0.001 for the \textit{spatial} condition, p<0.01 for the two others).
Since in all three conditions luminance is identical in episodes 1 and 2, luminance cannot explain the observed differences.

Second, we checked whether 
the differences in the pupil traces could be explained by the novelty of a state during episode 1, or familiarity with the state in episode 2 \citep{Otero11}, rather than by reward-based learning.
In a further control experiment,
a different set of participants saw a sequence of states, replayed from the main experiment.
In order to ensure that participants were focusing on the state sequence and engaged in the task, they had to push a button in each state (freely choosing either 'a' or 'b'), and count the number of states from start to goal.
Stimuli, timing and data analysis were the same as in the main experiment.
The strong difference after $1000\,ms$ in state D2, that we observed in Fig.~\ref{fig:PupilDiameter_GrandAvg}[b], was absent in the control experiments (Fig.~\ref{fig:Control_PupilDiameter})
indicating that the significant differences in pupil dilation in response to state D2 cannot be explained by novelty or familiarity alone. The findings in the control experiment also exclude other interpretations of correlations of pupil diameter such as memory formation in the absence of reward.

In summary,  across three different stimulus modalities, the single reward received at the end of the first episode  strongly influenced the pupil responses to the same stimuli later in episode 2.
Importantly, this effect was observed not only in state D1 (one step before the goal)
but also in state D2 (two steps before the goal). 
Furthermore, a mere engagement in button presses while observing a sequence of stimuli, as in the control experiment,
did not evoke the same pupil responses as the main task.
Together these results suggested
that the single reward at the end of  the first episode triggered 
increases in pupil diameter during later encounters of the same state.
The increases observed in state D1 are consistent with an interpretation that pupil diameter reflects state value $V$, action value $Q$, or TD error -
but do not inform us whether $Q$-value, $V$-value, or TD-error are estimated by the brain using RL with or without eligibility trace.
However, 
the fact that very similar changes are also observed in state D2 excludes the possibility that the learning-related contribution to the pupil diameter
can be predicted by RL without eligibility trace.

While our experiment was not designed to identify whether the pupil response reflects TD-errors or state values, we tried to address this question based on a model-driven analysis of the pupil traces.
First, 
we extracted all pupil responses after the onset of non-goal states and calculated the TD-error (according to the best-fitting model, \textit{Q-$\lambda$}, see next section) of the corresponding state transition.
We found that the pupil dilation was much larger after transitions with high TD-error compared to transitions with zero TD-error (Fig.~\ref{fig:RPE_Regression_Raw}[a] and Methods).
Importantly,
these temporal profiles of the pupil responses to states with high TD-error had
striking similarities across the three experimental conditions,
whereas the mean response time course 
was different across the three conditions (Fig.~\ref{fig:RPE_Regression_Raw}[c]).
This suggests that the underlying physiological process causing the TD-error-driven component in the
pupil responses was invariant to stimulation details.
Second, a statistical analysis including data with low, medium, and high TD-error confirmed the correlation of pupil dilation with TD error (Fig.~\ref{fig:RegressionAndPermTest}).
Third, a further qualitative analysis revealed that TD-error, rather than value itself, was a factor modulating pupil dilation (Fig.~\ref{fig:RPE_Regression_Raw}[b]).

\subsection{Estimation of the time scale of the behavioral eligibility trace using Reinforcement Learning Models}

Given the behavioral and physiological evidence for RL 'with eligibility trace', we wondered whether our findings are consistent with earlier studies \citep{Daw11, Tartaglia17, Bogacz07b} where  several variants of reinforcement learning algorithms were fitted to the experimental data.
We considered algorithms with and (for comparison) without eligibility trace.
Eligibility traces $e_n(s,a)$ can be modeled as a memory of past state-action pairs $(s,a)$ in an episode.
At the beginning of each episode all twelve eligibility trace values 
(two actions for each of the six decision states) were set to $e_n(s,a)=0$.
At each discrete time step $n$, the eligibility of the current state-action pair was set to 1,
while that of all others decayed by a factor $\gamma\lambda$ according to \citep{Singh96}
\begin{align}
 	e_{n}(s,a) &= \left\{\begin{array} {l l}
	1 &\text{if } s = s_n, a = a_n  \\
	\gamma \lambda e_{n-1}(s,a) &\text{otherwise}.\\
	\end{array}\right.
	\label{eq:ET_lambda}
\end{align}
The parameter $\gamma \in [0,1]$ exponentially discounts a distal reward, as commonly described in neuroeconomics \citep{Glimcher13} and machine learning \cite{Sutton18}; the parameter $\lambda \in [0,1]$ is called the decay factor of the eligibility trace.
The limit case $\lambda =0$ is interpreted as no memory and represents an instance of \textit{RL without eligibility trace}.
Even though the two parameters $\gamma$ and $\lambda$
appear as a product in equation \ref{eq:ET_lambda} so that the decay of the eligibility trace depends on both,
they have different effects in spreading the reward information
from one state to the next (cf. Eq.~\ref{eq:QLearnQupd} in Methods).
After many trials 
the $V$-values of states, or $Q$-values of actions, approach final
values which only depend on $\gamma$, but not on $\lambda$.
Given a parameter $\gamma>0$, the choice of $\lambda$ determines
how far value information spreads in a single trial.
Note that for $\lambda =0$ (\textit{RL without eligibility trace}),  Eq.~\ref{eq:ET_lambda} assigns an eligibility $e_n=1$ to state $D1$ in the first episode at the moment of the transition to the goal (while the eligibility at state $D2$ is $0$).
These values of eligibility traces lead to a spread of reward information
from the goal to state D1, but not to D2, at the end of the first episode in models without eligibilty trace
(cf. Eq.~\ref{eq:QLearnQupd} and Fig.~\ref{fig:PSA_ModelPrediction} in Methods), hence
the qualitative argument for episodes one and two as sketched in Fig.~\ref{fig:F1_TaskAndHyp}.

We considered eight common algorithms to explain the behavioral data:
Four algorithms belonged to the class of RL with eligibility traces. 
The first two, \textit{SARSA-$\lambda$} and \textit{Q-$\lambda$}
(see Methods, Eq. \ref{eq:QLearnQupd})
implement a memory of past state-action pairs by
an eligibility trace as defined in Eq. \ref{eq:ET_lambda};
as a member of the Policy-Gradient family, we implemented a variant of \textit{Reinforce} \citep{Sutton18, Williams92}, which memorizes all state-action pairs of an episode.
A fourth algorithm with eligibility trace is the 3-step Q-learning algorithm
\citep{Sutton18, Watkins89,Mnih16}, 
which keeps memory of past states and actions over three steps (see Discussion and Methods).
From the model-based family of RL, we chose the \textit{Forward Learner} \cite{Glaescher10}, which memorizes not state-action pairs, but learns a state-action-next-state model, and uses it for offline updates of action-values.
The \textit{Hybrid Learner} \citep{Glaescher10}
combines the \textit{Forward Learner} with \textit{SARSA-0}.
As a control, we chose two algorithms belonging to the class of RL without eligibility traces (thus modeling the null hypothesis): \textit{SARSA-$0$} and \textit{Q-$0$}.

We found that the four RL algorithms with eligibility trace explained human behavior better than the
\textit{Hybrid Learner}, which was the top-scoring among all other RL algorithms.
Cross-validation confirmed that our ranking based on 
the Akaike Information Criterion (AIC,  \cite{Akaike74}; see Methods) was robust. 
According to the Wilcoxon rank-sum test,
the probability that the \textit{Hybrid Learner} ranks better than one of the three RL algorithms with explicit eligibility traces was below 14$\%$ in each of the conditions and below 0.1$\%$ for the aggregated data ($p<0.001$, Table \ref{tab:AIC_CV_pVals} and Methods). The models \textit{Q}-$\lambda$ and \textit{SARSA}-$\lambda$ with eligbility trace  
performed each significantly better than the corresponding models
 \textit{Q-$0$} and  \textit{SARSA-$0$} without eligbility trace.

\begin{figure}
\centering
\includegraphics[width=0.9\textwidth]{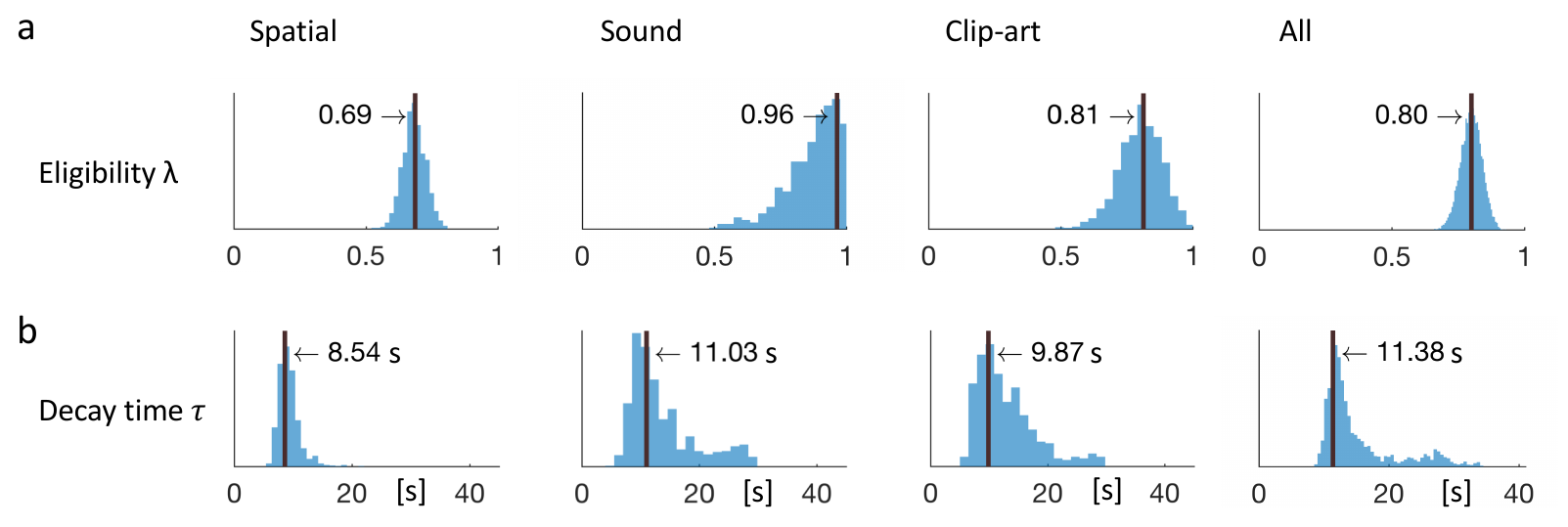}
\caption{\textbf{Eligibility for reinforcement decays with a time-scale $\tau$ in the order of 10 seconds.}
  The behavioral data of each experimental condition constrain the free parameters of the model \textit{Q-$\lambda$} to the ranges indicated by the blue histograms (see Methods and Fig.~\ref{fig:QLearn_MCMC_posterior})
  	 \textbf{[a]}
	 Distribution over the eligibility trace parameter $\lambda$ in Eq. \ref{eq:ET_lambda} (discrete time steps).
	Vertical black lines indicate the values that best explain the data (maximum likelihood, see Methods).
	All values are significantly different from zero.
	\textbf{[b]}
	Modeling eligibility in continuous time with
        a time-dependent decay (Methods, Eq. \ref{eq:ET_tau}), instead of a discrete per-step decay.
	The behavioral data constrains the time-scale parameter $\tau$ to around 10 seconds.
	Values in the column \textit{All} are obtained by fitting $\lambda$ and $\tau$ to the aggregated data of all conditions.
		}
\label{fig:MCMC_Posterior_ETonly}
\end{figure}

Since the ranks of the four RL algorithms with eligibility traces were not significantly different, we focused on one of these, viz. \textit{Q-$\lambda$}. We
wondered whether the parameter $\lambda$ that characterizes the decay of the eligibility trace in  Eq. \ref{eq:ET_lambda}
could be linked to a time scale.
To answer this question, we proceeded in two steps.
First, we analyzed the human behavior in discrete
time steps corresponding to state transitions.
We found that the best fitting values (maximum likelihood, see Methods) of the eligibility trace parameter $\lambda$  were 0.81 in the \textit{clip-art}, 0.96 in the \textit{sound}, and 0.69 in the \textit{spatial} condition (see Fig.~\ref{fig:MCMC_Posterior_ETonly}).
These values 
are all significantly larger than zero (p<0.001) indicating the presence of an eligibility trace consistent with our findings in the previous subsections.

In a second step,
we modeled the same action sequence in continuous time, taking into account the measured inter-stimulus interval which was the sum of the reaction time plus a random delay of $2.5$ to $4$ seconds after the push-buttons was pressed.
The reaction times were similar in the \textit{spatial}- and \textit{clip-art} condition, and slightly longer in the \textit{sound} condition with the following $10\%$, $50\%$ and $90\%$ percentiles: Spatial: $[0.40, 1.19, 2.73]$, Clip-Art: $[0.50, 1.11, 2.57]$,  Sound: $[0.67,  1.45,  3.78]$ seconds.
In this continuous-time version of the eligibility trace model, both the discount factor $\gamma$ and the decay factor $\lambda$ were integrated into a single time constant $\tau$ that describes the decay of the memory of past state-action associations in continuous time.
We found maximum likelihood values for $\tau$ around 10 seconds  (Fig.~\ref{fig:MCMC_Posterior_ETonly}), corresponding to 2 to 3 inter-stimulus intervals.
This implies that an action taken 10 seconds before a reward was reinforced and
associated with the state in which it was taken --  even if one or several decisions happened in between (see Discussion). 

Thus eligibility traces, i.e. memories of past state-action pairs, decay over about 10 seconds and
can be linked to a reward occurring during that time span.

\begin{table}
\centering
\begin{tabular}{c}
\includegraphics[width=0.95\textwidth]{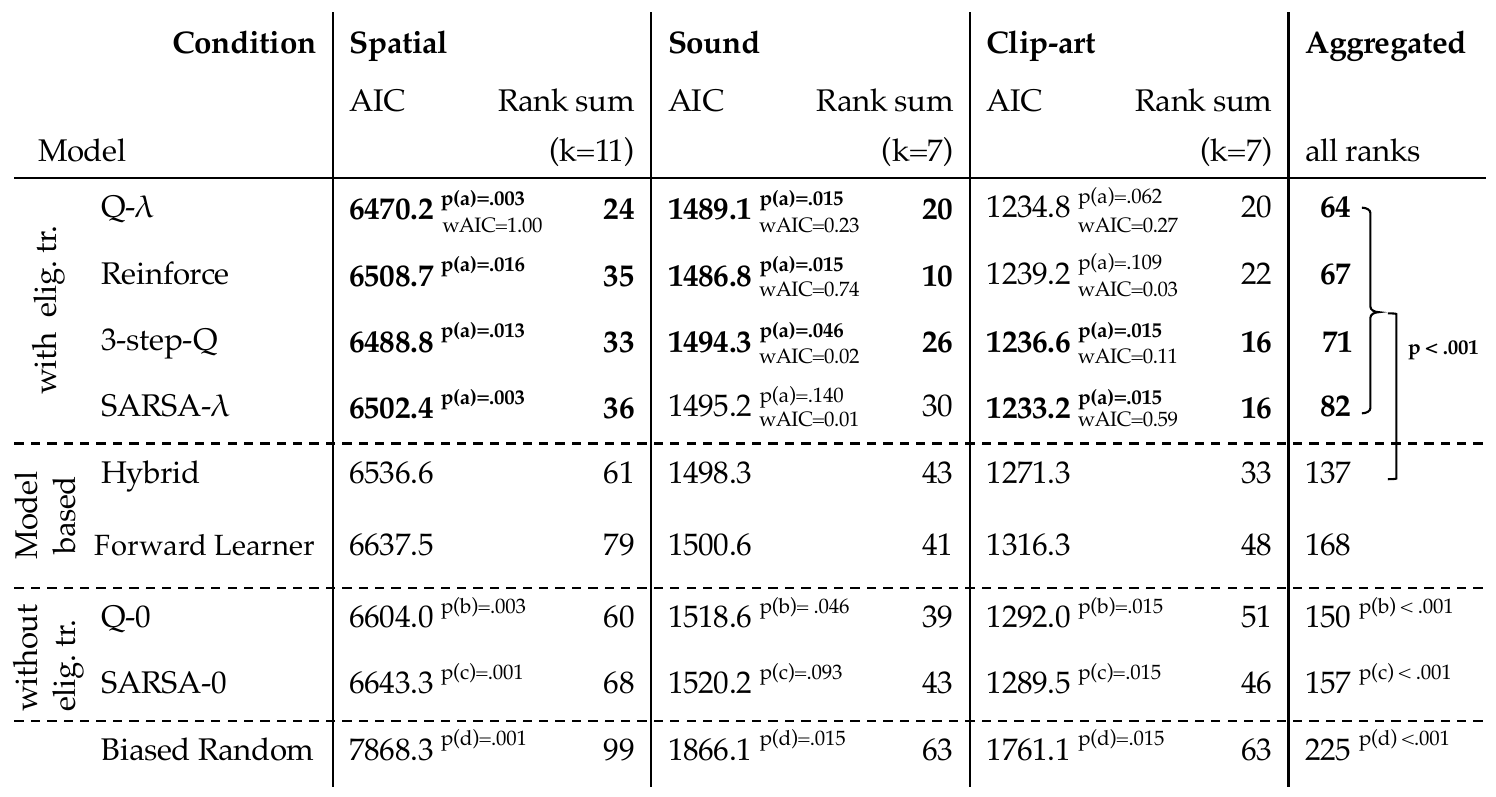}
\end{tabular}
\caption{ \textbf{Models with eligibility trace explain behavior significantly better than alternative models}. 
  Four reinforcement learning models with eligibility trace (Q-$\lambda$, REINFORCE, SARSA-$\lambda$, 3-step-Q), two model-based algorithms  (Hybrid, Forward Learner), two RL models without eligibility trace (Q-0, SARSA-0), and a null-model (Biased Random, Methods) were  fitted to the human behavior, separately for each experimental condition (spatial, sound, clip-art).
  Models with eligibility trace ranked higher than those without 
  (lower Akaike Information Criterion, AIC, evaluated on all participants performing the condition).
  $wAIC$ indicates the \textit{normalized Akaike weights} \citep{Burnham04}, values <0.01 are not added to the table. Note that only models \textit{with eligibility trace} have  $wAIC > 0.01$.
 The ranking is stable as indicated by the sum of $k$ rankings (column \textit{rank sum})
on test data, in $k$-fold crossvalidation (Methods).
P-values refer to the following comparisons:
P(a): Each model in the \textit{with eligibility trace} group was compared with the best model \textit{without eligibility trace} (Hybrid in all conditions); models for which the comparison is significant are shown in bold.
P(b): \textit{Q-0} compared with \textit{Q-$\lambda$}.
P(c): \textit{SARSA-0} compared with \textit{SARSA-$\lambda$}.
P(d): \textit{Biased Random} compared with the second last model, which is \textit{Forward Learner} in the clip-art condition and \textit{SARSA-0} in the two others.
In the \textbf{Aggregated} column, we compare the same pairs of models, taking into account all ranks across the three conditions.
All algorithms with eligibility trace explain the human behavior better than algorithms without eligibility trace.
Differences among the four models with eligibility trace are not significant.
In each comparison, $k$ pairs of individual ranks are used to compare pairs of models and obtain the indicated p-values (Wilcoxon rank-sum test, Methods).
} \label{tab:AIC_CV_pVals}
\end{table}

\section{Discussion} 

Eligibility traces provide a mechanism for learning temporally extended action sequences from a single reward (one-shot). 
While one-shot learning is a well-known phenomenon for tasks such as image recognition
\cite{Standing73,Brady08} and one-step decision making \cite{Duncan16, Greve17, Rouhani18}
it has so far not been linked to Reinforcement Learning (RL) with eligibility traces in multi-step decision making.

In this study, we asked whether humans use eligibility traces when learning long sequences from delayed feedback.
We formulated mutually exclusive hypotheses, which predict directly observable changes in behavior and in physiological measures when learning with or without eligibility traces. 
Using a novel paradigm, we could reject the null hypothesis of learning without eligibility trace in favor of the alternative hypothesis of learning with eligibility trace.

Our multi-step decision task shares aspects
with earlier work in the neurosciences
\citep{Pessiglione06, Glaescher10, Daw11, Niv12, ODoherty17, Walsh11},
but overcomes their limitations (i) by using a recurrent graph structure of the environment
that enables relatively long episodes \citep{Tartaglia17},
 and (ii) by implementing an 'on-the-fly' assignment rule for state-action transitions during the first episodes.
This novel design allows the study of human learning in specific and controlled conditions, without interfering with the participant's free choices.

A difficulty in the study of eligibility traces, is that in the relatively simple tasks typically used in animal \citep{Pan05} or human \citep{Daw11, Tartaglia17, Walsh11, Bogacz07b, Weinberg12, Gureckis09} studies, the two hypotheses make qualitatively different predictions only during the first episodes: At the end of the first episode, algorithms in the class of RL without eligibility trace update only the value of state D1 (but not of D2. see Fig.~\ref{fig:F1_TaskAndHyp}, Null hypothesis).
Then, this value of D1 will drive learning at state D2 when the participants move from D2 to D1 during episode two. 
In contrast, algorithms in the class of RL with eligibility trace, update D2 already during episode one.
Therefore, only during episode two, the behavioral data permits a clean, qualitative dissociation between the two classes.
On the other hand, the fact that for most episodes, the differences are not qualitative, is the reason why eligibility trace contributions have typically been statistically inferred from many trials through model selection \citep{Daw11, Tartaglia17, Walsh11, Bogacz07b, Pan05, Gureckis09}.
Here, by a specific task design and a focus on episodes one and two, we provided directly observable, qualitative, evidence for learning with eligibility traces from behavior and pupil data without the need of model selection. 

In the quantitative analysis, RL models with eligibility trace explained the behavioral data significantly better than the best tested RL models without.
There are, however, in the reinforcement learning literature, several alternative algorithms that would also account for one-shot learning but  do not rely on the explicit eligibility traces formulated in Eq. \ref{eq:ET_lambda}.
First, $n$-step reinforcement learning algorithms \cite{Sutton18,Watkins89,Mnih16}
compare the value of a state not with that of its direct neighbor but of neighbors that are $n$ steps away. These algorithms are closely related to eligibility traces and in certain cases even mathematically equivalent \cite{Sutton18}.
Second, reinforcement learning algorithm with storage of past sequences \cite{Moore93,Blundell16,Mnih16}
 enable the offline replay of the first episode so as to update values of states far away from the goal. 
 While these approaches are formally different from eligibility traces, they nevertheless implement the idea of eligibility traces
as memory of past state-action pairs \cite{Crow68,Fremaux16}, albeit
in a different algorithmic framework.
For example, prioritized sweeping with small backups \cite{Seijen13} is an offline algorithm that is, if applied 
to our deterministic environment after the end of the first episode, equivalent to both episodic control \cite{Brea17a}  and an eligibility trace.
Interestingly, the two model-based algorithms (\textit{Forward Learner} and \textit{Hybrid}) would in principle be able to explain one-shot learning since reward information is spread, after the first episode, throughout the model, via offline Q-value updates. Nevertheless, when behavioral data from our experiments were fitted across all 7 episodes, the two model-based algorithms performed significantly worse than the RL models with explicit eligibility traces.
Since our experimental design does not allow us to distinguish between these different algorithmic implementations of closely related ideas, we put them all in the class of RL with eligibility traces.

Importantly, RL algorithms with explicit eligibility traces
\citep{Sutton88a,Williams92, Peng96,Fremaux16,Izhikevich07} 
can be mapped to known synaptic and circuit mechanisms
\citep{Yagishita14, He15, Bittner17, Fisher17, Gerstner18}.
A time scale of the eligibility trace of about 10 seconds in our experiments is in the range of, but a bit longer than those observed for dopamine modulated plasticity in the striatum \citep{Yagishita14}, 
serotonin and norepinephrine modulated plasticity in the cortex \citep{He15}, 
or complex-spike plasticity in hippocampus \citep{Bittner17}, 
but shorter than the  time scales of minutes reported in hippocampus \citep{Brzosko17}.
The basic idea for the relation of eligibility traces as in Eq. \ref{eq:ET_lambda} to 
experiments on synaptic plasticity is that choosing
action $a$ in state $s$ leads to co-activation of neurons and leaves a trace at the synapses connecting
those neurons. 
A later phasic neuromodulator signal will transform the trace into a change of the synapses so that taking action $a$ in state $s$ becomes more likely in the future \cite{Sutton18, Gerstner18, Crow68, Izhikevich07}.
Neuromodulator signals could include dopamine \cite{Schultz15}, but reward-related signals could also be conveyed, together with novelty or attention-related signals,  by other modulators \citep{Fremaux16}.

Since in our paradigm the ISI was not systematically varied, we cannot distinguish between an eligibility trace with purely time-dependent, exponential decay, and one that decays discretely, triggered by events such as states or actions. 
Future research needs to show whether the decay is event-triggered or defined by molecular characteristics, independent of the experimental paradigm.

Our finding that changes of pupil dilation correlate with reward-driven variables of reinforcement learning (such as value or TD error)
goes beyond the changes linked to state recognition reported earlier \cite{Otero11,Kucewicz18}.
Also, since non-luminance related pupil diameter is  influenced by the neuromodulator norepinephrine \cite{Joshi16}
while reward-based learning is associated with the neuromodulator dopamine \cite{Schultz15}, our findings suggest that the roles, and regions of influence, of neuromodulators could be mixed \cite{Berke18,Fremaux16} and less well segregated than suggested by earlier theories.

\begin{figure*}
	\centering
	\includegraphics[width=\textwidth]{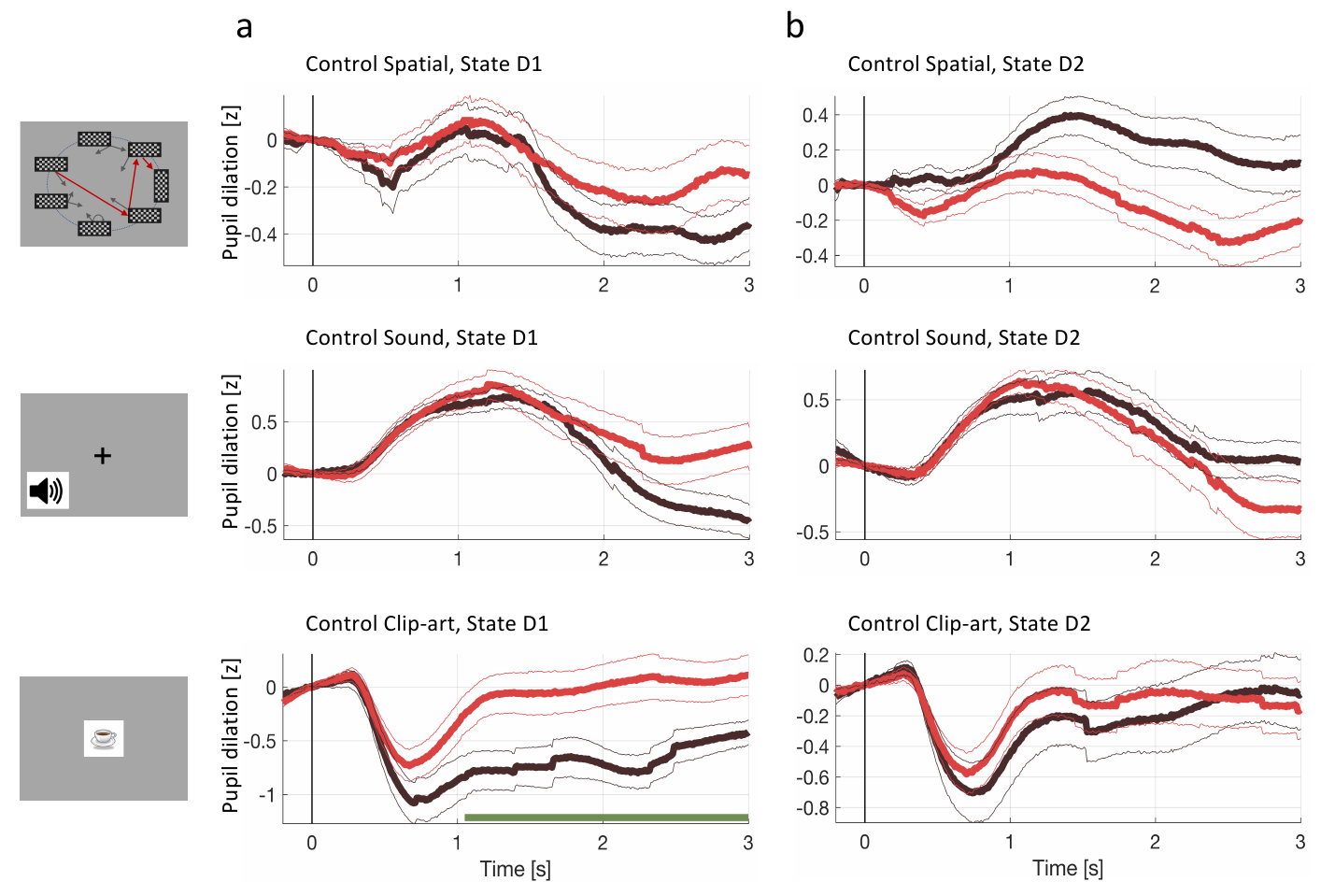}
	\caption{\textbf{Pupil dilation during the second control experiment.} In the second control experiment, different participants passively observed state sequences which were recorded during the main experiment. Data analysis was the same as for the main experiment. \textbf{[a]} Pupil time course after state onset ($t=0$) of state D1 (before goal). \textbf{[b]} State D2 (two before goal).
Black traces show the pupil dilation during episode one, red traces during episode two.
At state D1 in the \textit{clip-art} condition the pupil time course shows a separation similar to the one observed in the main experiment. This suggest that participants may recognize the clip-art image that appears just before the final image. Importantly in state D2, the pupil time course during episode two is qualitatively different from the one in the main experiment (Fig.~\ref{fig:PupilDiameter_GrandAvg}).}
	\label{fig:Control_PupilDiameter}
\end{figure*}

From the qualitative analysis of the pupillometric data of the main experiment (Fig.~\ref{fig:PupilDiameter_GrandAvg}), together with those of the control experiment (Fig.~\ref{fig:Control_PupilDiameter}),
we concluded that changes in pupil dilation 
reflected a learned, reward-related property of the state.
In the context of decision making and learning, pupil dilation is most frequently associated with violation of an expectation in the form of a reward prediction error or stimulus prediction error as in an oddball-task \citep{Nieuwenhuis11}.
However, 
our experimental paradigm was not designed to decide whether pupil diameter correlates stronger with state values  or TD-errors.
Nevertheless, a more systematic  analysis (see Methods and
Fig.~\ref{fig:RPE_Regression_Raw}) suggests that correlation of pupil dilation with TD-errors is stronger than correlation with state values.

\subsection{Conclusion} 
Eligibility traces are a fundamental factor underlying the human capability of quick learning and adaptation.
They implement a memory of past state-action associations and are a crucial element to efficiently solve the credit assignment problem in complex tasks \citep{Sutton18, Gerstner18, Izhikevich07}.
The present study provides both qualitative and quantitative evidence for one-shot sequence-learning with eligibility traces.
The correlation of the pupillometric signals with an RL algorithm with eligibility traces suggests
that humans not only exploit memories of past state-action pairs in behavior
but also assign reward-related values to these memories.
The consistency and similarity of our findings across three experimental conditions suggests that the underlying cognitive, or neuromodulatory, processes are independent of the stimulus modality.
It is an interesting question for future research to actually identify the neural implementation of these memory traces.

\clearpage
 
\beginsupplement
\section{Materials and Methods (Supplementary) }

\subsection{Experimental conditions}
We implemented three different experimental conditions based on the same Markov Decision Process (MDP) of Fig.~\ref{fig:F2_Task_Cond_Behav}[a].
The conditions only differed in the way the states were presented to the participant.
Furthermore, in order to collect enough samples from early trials, where the learning effects are strongest, participants did not perform one long experiment. 
Instead, after completing seven episodes in the same environment, the experiment paused for 45 seconds while participants were instructed to close and relax their eyes.
Then the experiment restarted with a new environment: the transition graph was reset, a different, unused, stimulus was assigned to each state, and the participant had to explore and learn the new environment.
We instructed the participants to reach the goal state as often as possible within a limited time (12 minutes in the \textit{sound} and  \textit{clip-art} condition, 20 minutes in the \textit{spatial} condition).
On average, they completed $48.1$ episodes ($6.9$ environments) in the \textit{spatial} condition  , $19.4$ episodes ($2.7$ environments) in the \textit{sound} condition and $25.1$ episodes ($3.6$ environments) in the \textit{clip-art} condition

In the \textit{spatial} condition, each state was defined by the location (on an invisible circle) on the screen of a 100x260 pixels checkerboard image, flashed for 100ms, (Fig.~\ref{fig:F2_Task_Cond_Behav}[d]).
The goal state was represented by the same rectangular checkerboard, but rotated by 90 degrees.
The checkerboard had the same average luminance as the grey background screen.
In each new environment, the states were randomly assigned to locations and 
the checkerboards were rotated (states: 260x100 pixels checkerboard, goal: 100x260).

In the \textit{sound} condition each state was represented by a unique acoustic stimulus (tones and natural sounds) of $300ms$ to $600ms$ duration.
New, randomly chosen, stimuli were used in each environment.
At the goal state an applause was played.
An experimental advantage of the \textit{sound} condition is that a change in the pupil dilation cannot stem from a luminance change but must be due to a task-specific condition.

In the \textit{clip-art} condition, each state was represented by a unique 100x100 pixel clip-art image that appeared for $300ms$ in the center of the screen.
For each environment, a new set of images was used, except for the goal state which was always the same (a person holding a trophy) in all experiments.

The screen resolution was 1920x1080 pixels.
In all three conditions, the background screen was grey with a fixation cross in the center of the screen.
It was rotated from $+$ to $\times$ to signal to the participants when to enter their decision by pressing one of two push-buttons (one in the left and the other in the right hand).
No lower or upper bound was imposed on the reaction time. 
The next state appeared after a random delay of $2.5$ to $4$ seconds after the push-buttons was pressed.
Prior to the actual learning task, they performed a few trials to check they all understood the instructions.
While the participants performed the \textit{sound}- and \textit{clip-art} conditions, we recorded the pupil
diameter using an SMI iViewX high speed video-based eye tracker (recorded at $500 Hz$, down-sampled to $100Hz$ for the analysis by averaging over 5 samples).
From participants performing the \textit{spatial} condition, we recorded the pupil diameter using a $60Hz$ Tobii Pro tracker.
An eye tracker calibration protocol was run for each participant.
All experiments were implemented using the Psychophysics Toolbox \citep{Brainard97}.

The number of participants performing the task was: \textit{sound}: 15; \textit{clip-art}: 12; \textit{spatial}: 22 participants; Control \textit{sound}: 9; Control \textit{clip-art}: 10; Control \textit{spatial}: 12.  
The participants were recruited from the EPFL students pool. 
They had normal or corrected-to-normal vision.
Experiments were conducted in accordance with the Helsinki declaration and approved by the ethics commission of the Canton de Vaud (164/14 Titre: Aspects fondamentaux de la reconnaissance des objets : protocole g\'{e}n\'{e}ral). All participants were informed about the general purpose of the experiment and provided written, informed consent. They were told that they could quit the experiment at any time they wish.

\subsection{Pupil data processing}
\begin{figure*}
	\centering
	\includegraphics[width=0.8\textwidth]{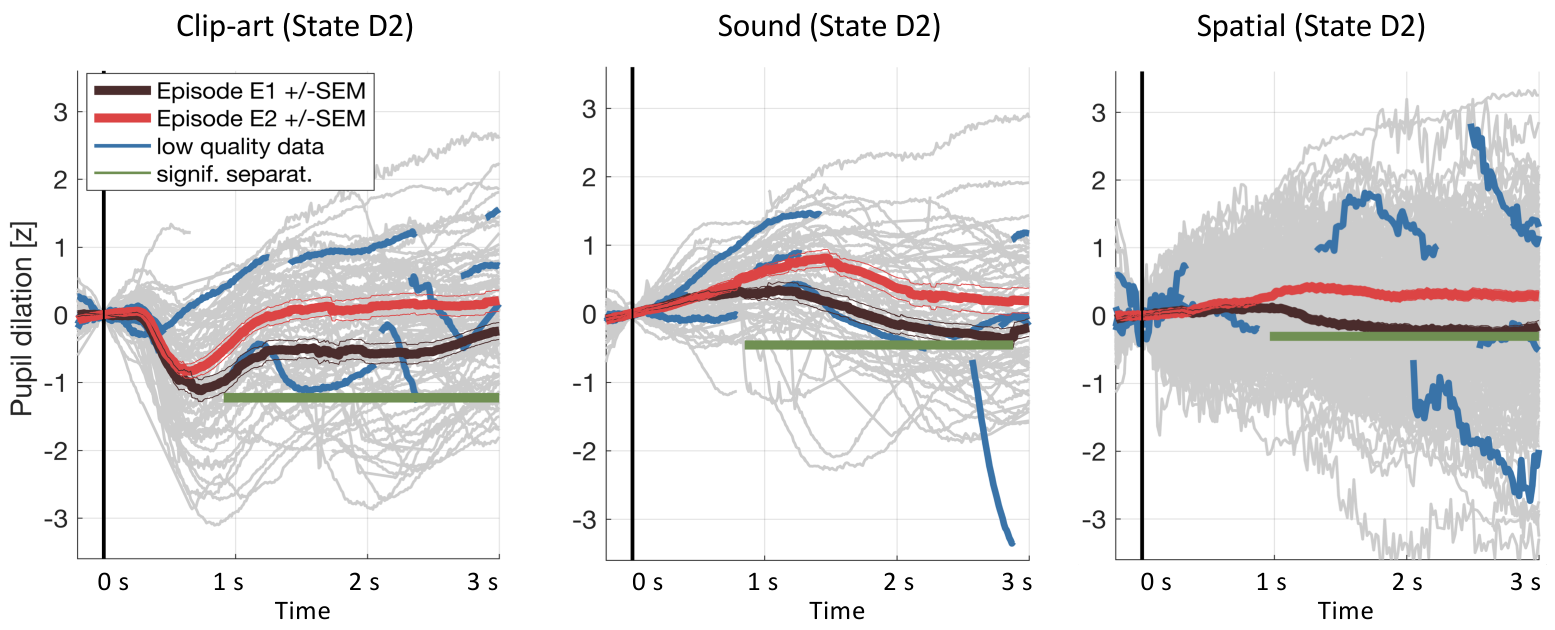}
	\caption{
	\textbf{Results including low-quality pupil traces.}
          	We repeated the pupil data analysis at the crucial state $D2$ including all data (including traces with less than $50\%$ of data within the $3s$ window and with z-values outside $\pm3$). 
          	Gray curves in the background show all recorded pupil traces.
          	The highlighted blue curves show a few, randomly selected, low-quality pupil traces.
          	Including these traces does not affect the result.
	 }
	\label{fig:Pupil_low_quality_traces}
\end{figure*}

Our data processing pipeline followed recommendations described in \citep{Mathot17}.
Eye blinks (including $100ms$ before, and $150ms$ after) were removed and short blocks without data (up to $500ms$) were linearly interpolated.
In all experiments, participants were looking at a fixation cross which reduces artifactual pupil-size changes \citep{Mathot17}.
For each environment, the time-series of the pupil diameter during the 7 episodes was extracted and then normalized to zero-mean, unit variance.
This step renders the measurements comparable across participants and environments.
We then extracted the pupil recordings at each state from $200ms$ before to $3000ms$ after each state onset and applied subtractive baseline correction where the baseline was taken as the mean in the interval [$-100ms$, $+100ms$].
Taking the $+100ms$ into account does not interfere with event-specific effects because they develop only later (>220ms according to \cite{Mathot17}), but a symmetric baseline reduces small biases when different traces have different slopes around t=0ms.
We considered event-locked pupil responses with z-values outside $\pm 3$ as outliers and excluded them from the main analysis.
We also  excluded pupil traces with less than $50\%$ eye-tracker data within the time window of interest, because very short data fragments do not provide information about the characteristic time course of the pupil trace after stimulus onset.
As a control, Figure ~\ref{fig:Pupil_low_quality_traces} shows that the conclusions of our study are not affected if we drop the two conditions and include all data.

\subsection{Action assignment in the Markov Decision Process}

Actions in the graph of Fig.~\ref{fig:F2_Task_Cond_Behav} were assigned to transitions during the first few actions as explained in the main text.
However, our learning experiment would become corrupted if participants would discover that in the first episode any three actions lead to the goal.
First, such knowledge would bypass the need to actually learn state-action associations, and second, the knowledge of "distance-to-goal" implicitly provides reward information even before seeing the goal state.
We avoided the learning of the latent structure by two manipulations:
First, if in episode one of a new environment a participant repeated the exact same action sequence as in the previous environment, or if they tried trivial action sequences (a-a-a or b-b-b), the assignment of the third action led from state D1 to Z, rather than to the Goal.
This was the case in about $1/3$ of the first episodes (\textit{spatial}: $48/173$, \textit{sound}: $20/53$ \textit{clip-art}: $23/49$).
The manipulation further implied that participants had to make decisions against their potential left/right bias.
Second, an additional state H (not shown in Fig.~\ref{fig:F2_Task_Cond_Behav}) was added in episode one in some environments (\textit{spatial}: $23/173$, \textit{sound}: $6/53$ \textit{clip-art}: $8/49$).
Participants then started from H (always leading to S) and the path length to goal was four steps.
Interviews after the experiment showed that no participant became aware of the experimental manipulation and, importantly, they did not notice that they could reach the goal with a random action sequence in episode one.

\subsection{Reinforcement Learning Models}

For the RL  algorithm $Q-\lambda$ (see Algorithm~\ref{Alg:Qlearning}), four quantities are important: the reward $r$;
the value $Q(s,a)$
of a state-action association such as taking action 'b' in state D2; 
the value $V(s)$ of the state itself, defined as
the larger of the  two $Q$-values in that state, i.e., $V(s) = {\rm max}_{\tilde{a}} Q(s,\tilde{a})$; 
and the TD-error  (also called Reward Prediction Error or RPE) calculated at the end of the $n^{th}$ action
after the transition from state $s_n$ to  $s_{n+1}$
\begin{equation}
 {\rm RPE}(n \rightarrow n+1) = r_{n+1} + \gamma \cdot \, V(s_{n+1}) - Q(s_{n}, a_{n}) \\
	\label{eq:RewardPredictionError}
\end{equation}
Here $\gamma$ is the discount factor and $V(s)$ is the estimate of the  discounted future reward that can maximally be collected when starting from state $s$. 
Note that RPE is different from reward. In our environment a reward occurs only at the transition from state D1 to state G whereas reward prediction errors occur in episodes 2 - 7 also several steps before the reward location is reached.

The table of values $Q(s,a)$ is initialized at the beginning of an experiment
and then updated by combining the RPE and the 
eligibility traces $e_n(s, a)$ defined in the main text (Eq. \ref{eq:ET_lambda}),
\begin{equation}
Q(s, a) \leftarrow Q(s, a) + \alpha \cdot RPE(n) \cdot e_n(s, a) \, ,
\label{eq:QLearnQupd}
\end{equation}
where $\alpha$ is the learning rate.
Note that \textit{all} Q-values are updated, but changes in $Q(s_n,a_n)$ are proportional to the eligibility of the state-action pair $e_n(s, a)$.
In the literature the table $Q(s, a)$ is often initialized with zero, but 
since some participants pressed the left (or right) button more often than the other one, we identified for each participant the preferred action $a_{pref}$ and initialized $Q(s, a_{pref})$ with a small bias $b$, adapted to the data.

Action selection exploits the Q-values of Eq. \ref{eq:QLearnQupd} using a softmax criterion with temperature $T$:
\begin{align}
	 p(s,a) = \frac{exp(Q(s,a)/ T )}{\sum_{\tilde{a}} exp(Q(s,\tilde{a})/ T )} \, ,
	 \label{eq:psa_softmax}
\end{align}

As an alternative to the eligibility trace defined in Eq. \ref{eq:ET_lambda}, where the eligibility decays at each discrete time-step, we also modeled a decay in continuous time, defined as
\begin{equation}
	 e_t(s,a) = exp \left(-\frac{t - B(s,a)}{\tau} \right) \quad  {\rm if}  \; t>B(s,a)
	 \label{eq:ET_tau}
\end{equation}
and zero otherwise.
Here, $t$ is the time stamp of the current discrete step,
and $B(s,a)$ is the time stamp of the last time a state-action pair $(s,a)$ has been selected.
The discount factor $\gamma$ in Eq. \ref{eq:RewardPredictionError} is kept, while in Eq. \ref{eq:ET_tau} a potential discounting is absorbed into the single parameter $\tau$.

Our implementation of \textit{Reinforce} followed the pseudo-code of \textit{REINFORCE: Monte-Carlo Policy-Gradient Control (without baseline)} (\cite{Sutton18}, Chapter 13.3) which updates the action-selection probabilities at the end of each episode.
This requires the algorithm to keep a (non-decaying) memory of the complete state-action history of each episode.
We refer to \citep{Sutton18}, \citep{Glaescher10} and \citep{Peng96} for the pseudo-code and in-depth discussions of all algorithms.

\subsection{Parameter Fit and Model Selection}
The main goal of this study was to test the null-hypothesis 'RL without eligibility traces' from the behavioral responses at states D1 and D2 (Figs.  \ref{fig:F2_Task_Cond_Behav} [e] and [f]). By the design of the experiment, we collected relatively many data points from the early phase of learning, but only relatively few episodes in total. This contrasts with other RL studies, where participants typically perform longer experiments with hundreds of trials. As a result, the behavioral data we collected from each single participant is not sufficient to reliably extract the values of the model-parameters on a participant-by-participant basis. To find the most likely values of model parameters, we therefore pooled the behavioral recordings of all participants into one data set D.

\begin{figure*}
\centering
\includegraphics[width=\textwidth]{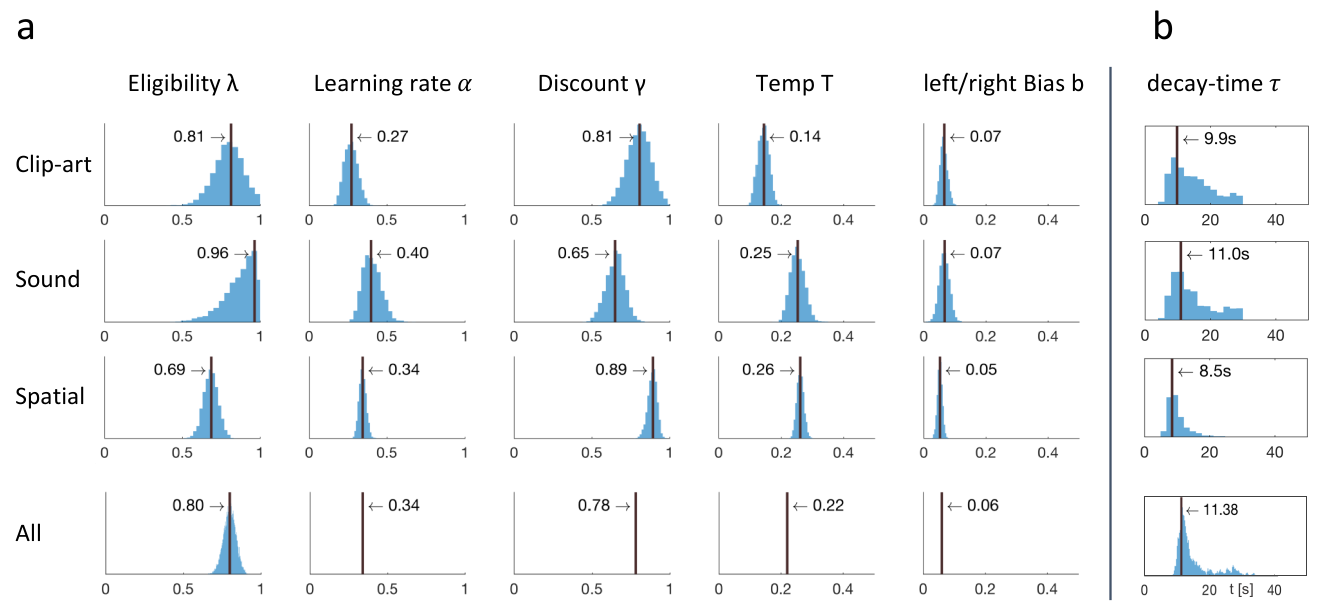}
\caption{ \textbf{Fitting results: behavioral data constrained the free parameters of \textit{Q-$\lambda$}.}
\textbf{[a]} For each experimental condition a distribution over the five free parameters is estimated by sampling.
The blue histograms show the approximate conditional posterior for each parameter (see methods).
Vertical black lines indicate the values of the 5-parameter sample that best explains the data (maximum likelihood, ML).
The bottom row (All) shows the distribution over $\lambda$ when fitted to the aggregated data of all conditions, with other parameters fixed to the indicated value (mean over the three conditions).
\textbf{[b]} Estimation of a time dependent decay ($\tau$ instead of  $\lambda$) as defined in equation \ref{eq:ET_tau}.
}
\label{fig:QLearn_MCMC_posterior}
\end{figure*}

Each learning model $m$ is characterized by a set of parameters $\theta^m = [\theta^m_1, \theta^m_2, ...]$.
For example, our implementation of the \textit{Q-$\lambda$} algorithm has five free parameters:
the eligibility trace decay $\lambda$;
the learning rate $\alpha$;
the discount rate $\gamma$;
the softmax temperature $T$;
and the bias $b$ for the preferred action.
For each model $m$, we were interested in the posterior distribution $P(\theta^m | D)$ over the free parameters $\theta^m$, conditioned on the behavioral data of all participants $D$.
This distribution was approximated by sampling using the Metropolis-Hastings Markov Chain Monte Carlo (MCMC) algorithm \citep{Hastings70}.
For sampling, MCMC requires a function $f(\theta^m, D)$ which is proportional to $P(\theta^m | D)$.
Choosing a uniform prior $P(\theta^m) = const$, and exploiting that $P(D)$ is independent of $\theta^m$, we can directly use the model likelihood $P(D | \theta^m)$:
\begin{align}
	 P( \theta^m | D) = \frac{P(D | \theta^m) P(\theta^m)}{P(D)} \propto P(D | \theta^m) := f(\theta^m, D).
	 \label{eq:MCMCposterior}
\end{align}
We calculated the likelihood $P(D | \theta^m)$ of the data as the joint probability of all action selection probabilities obtained by evaluating the model (Eqs. \ref{eq:ET_lambda}, \ref{eq:RewardPredictionError}, \ref{eq:QLearnQupd}, and \ref{eq:psa_softmax} in the case of $Q(\lambda)$) given a parameter sample $\theta^m$.
The log likelihood (LL) of the data under the model is
\begin{equation}
LL(D|\theta^m) =  \sum_{p=1}^N \, \sum_{j=1}^{E_p} \, \sum_{t=1}^{T_j}  log( p(a_t | s_t ; \theta^m)) \, ,
\end{equation}
where the sum is taken over all participants $p$, all environments $j$, and all actions $a_t$ a participant has taken in the environment $j$:

For each model, we collected $100'000$ parameter samples
(burn-in: 1500;
keeping only every $10^{th}$ sample;
50 random start positions;
proposal density: Gaussian with $\sigma=0.004$ for temperature $T$ and bias $b$,  and $\sigma=0.008$ for all other parameters).
From the samples we chose the $\hat{\theta}^m$ which maximizes the log likelihood (LL), calculated the $AIC_m$ and ranked the models accordingly.
The $AIC_m$ of each model is shown in Table~\ref{tab:AIC_CV_pVals}, alongside with the Akaike weights $wAIC_m$. The latter can be interpreted as the probability that the model $m$ is the best model for the data \citep{Burnham04}.
Note that the parameter vector $\hat{\theta}^m$  could be found by a hill-climbing algorithm towards the optimum, but such an algorithm does not give any indication about the uncertainty.
Here we obtained an approximate conditional posterior distribution  $p(\theta^m_{i} | D, \hat{\theta}^m_{j \neq i})$  for each component $i$ of the parameter vector $\theta^m$ (cf. Fig.~\ref{fig:QLearn_MCMC_posterior}).
We estimated this posterior for a given parameter $i$ by selecting only the $1 \%$ of all samples falling into a small neighborhood: $\hat{\theta}^m_{j} - \epsilon^m_j \leq  \theta_{ j}  \leq \hat{\theta}^m_{j} + \epsilon^m_j , i \neq j$.
We determined $\epsilon^m_j$ such that along each dimension $j$, the same percentage of samples was kept (about 22\%) and the overall number of samples was 1000.

One problem using the AIC for model selection stems from the fact that there are considerable behavioral differences across participants and the AIC model selection might change for a different set of participants.
This is why we validated the model ranking using $k$-fold cross-validation.
The same procedure as before (fitting, then ranking according to AIC) was repeated $K$ times,  but now we used only a subset of participants (training set) to fit $\hat{\theta}^m_k$ and then calculated the $LL^m_{k}$ and the $AIC^m_{k}$ on the remaining participants (test set).
We created the $K$ folds such that each participant appears in exactly one test set and in $K-1$ training sets.
Also, we kept these splits fixed across models, and evaluated each model on the same split into training and test set.
In each fold $k$, the models were sorted with respect to $AIC^m_{k}$, yielding $K$ lists of ranks.
In order to evaluate whether the difference between two models is significant, we compared their ranking in each fold (Wilcoxon rank-sum test on K matched pairs, p-values shown in Table \ref{tab:AIC_CV_pVals}).
The cross-validation results were summarized by summing the $K$ ranks (Table \ref{tab:AIC_CV_pVals}).
The best rank sum a model could obtain is $K$, and is obtained if it achieved the first rank in each of the $K$ folds.

\subsection{ $Q-\lambda$ model predictions}

\begin{algorithm}
  \caption{ \textbf{Q-$\lambda$} (and related models): \\
	For SARSA-$\lambda$ we replace the expression  $\max_{\tilde{a}} Q(s_{n+1}, \tilde{a})$ in line 9 by $Q(s_{n+1}, a_{n+1})$  where $a_{n+1}$ is the action taken in the next state $s_{n+1}$.
	For Q-$0$ and SARSA-$0$ we set $\lambda$ to zero.
	}
  \label{Alg:Qlearning}
  \begin{algorithmic}[1]
  \State Algorithm parameters: learning rate $\alpha \in (0,1]$, discount factor $\gamma \in [0,1]$, eligibility trace decay factor $\lambda \in [0,1]$, temperature $T \in (0, \infty)$ of softmax policy $p$, bias $b \in [0,1]$ for preferred action $a_{pref} \in \mathcal{A}$.
  \State Initialize $Q(s, a)=0$ and $e(s,a)=0$, for all $s \in \mathcal{S}, a \in \mathcal{A}$ \newline
        For preferred action $a_{pref} \in \mathcal{A}$ set $Q(s, a_{pref})=b$
  \For {each episode}
    \State Initialize state $s_n \in \mathcal{S}$
    \State Initialize step $n=1$
    \While {$s_n$ is not terminal}
      \State Choose action $a_n \in \mathcal{A}$ from $s_n$ with softmax policy $p$ derived from $Q$
      \State Take action $a_n$, and observe $r_{n+1} \in \mathbb{R}$ and $s_{n+1}\in \mathcal{S}$
      \State $RPE(n \rightarrow n+1) \leftarrow r_{n+1} + \gamma \max_{\tilde{a}} Q(s_{n+1}, \tilde{a}) - Q(s_n, a_n)$
      \State $e_n(s_n,a_n) \leftarrow 1$ 
      \For {all $s\in \mathcal{S}, a\in \mathcal{A}$}
        \State $Q(s, a) \leftarrow Q(s, a) + \alpha RPE(n \rightarrow n+1) e_n(s,a)$
        \State $e_{n+1}(s,a) \leftarrow \gamma \lambda e_n(s,a)$ 
      \EndFor
  \State $n \leftarrow n+1$
  \EndWhile
 \EndFor
\end{algorithmic}
\end{algorithm}

\begin{figure*}
	\centering
	\includegraphics[width=0.9\textwidth]{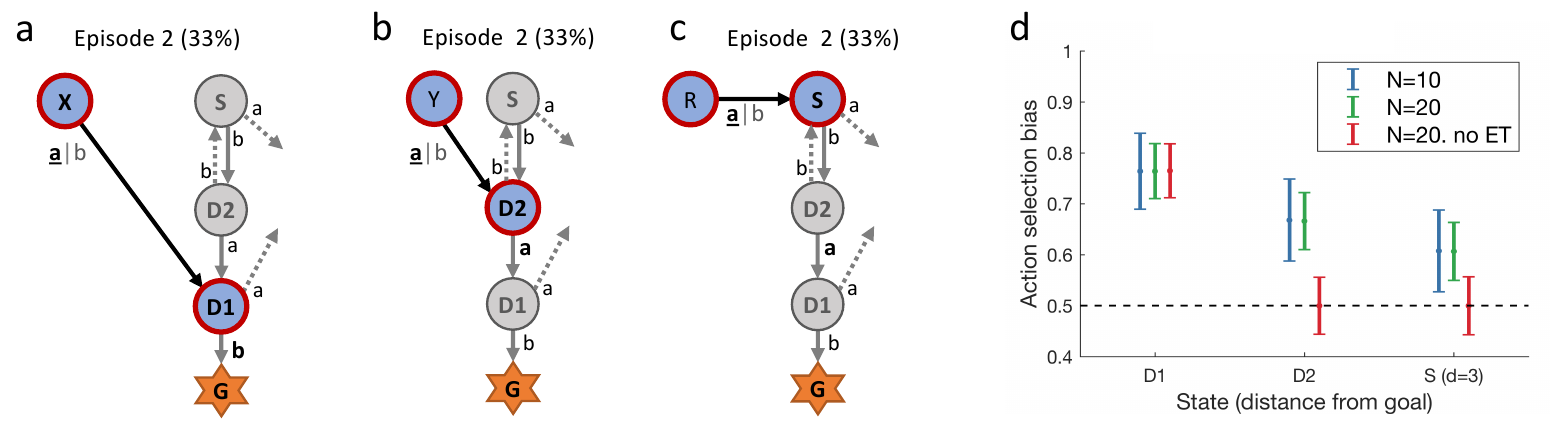}
	\caption{\textbf{Simulated experiment.}
( $Q-\lambda$ model).
          	\textbf{[a]} and \textbf{[b]:} Task structure (same as in Fig.~\ref{fig:F2_Task_Cond_Behav}). Simulated agents performed episodes one and two and we recorded the decisions at states $D1$ and $D2$ in episode two. \textbf{[c]:} Additionally, we also simulated the model's behavior at state $S$, by extending the structure of the (simulated) experiment with a new state $R$, leading to $S$.
          	\textbf{[d]:}
          	We calculated the action-selection bias at states $D1$, $D2$ and $S$ during episode two from the  behavior of $N=10$ (blue) and $N=20$ (green) simulated agents. 
          	The effect size (observed during episode two and visualized in panel [d]) decreases when (in episode one) the delay between taking the action and receiving the reward increases. It is thereby smallest at state $S$.
          	When setting the model's eligibility trace parameter $\lambda$ to $0$ (red, no ET), the effect at state $D1$ is not affected (see Eq. \ref{eq:ET_lambda}) while at $D2$ and $S$ the behavior was not reinforced.
          	Horizontal dashed line: chance level  $50\%$.
          	Errorbars: standard deviation of the simulated effect when estimating 1000 times the mean bias from $N=10$ and $N=20$ simulated agents with individually sampled model parameters.
		}
	\label{fig:PSA_ModelPrediction}
\end{figure*}

The $Q-\lambda$ model (see Algorithm~\ref{Alg:Qlearning}), and related models like $Sarsa-\lambda$, have previously been used to explain human data. 
We used those published results, in particular the parameter values from \citep{Glaescher10, Daw11, Tartaglia17}, to estimate the effect size, as well as the reliability of the result.
The published parameter values have a high variance: they differ across participants and across tasks. 
We therefore simulated different agents, each with its own parameters, sampled independently from a uniform distribution in the following ranges:
$\alpha \in [0.1, 0.5]$,  $\lambda \in [0.5, 1]$, $\gamma \in [0.5, 1]$, $T \in [0.125, 1]$ (corresponding to an inverse temperature $1/ T \in [1, 8]$), and $b=0$.
We then simulated episodes one and two of the experiment, applied the $Q-\lambda$ model to calculate the action-selection bias (Eq. \ref{eq:psa_softmax}) when the agents visit states $D1$,  $D2$ and also $S$ (see Fig.~\ref{fig:PSA_ModelPrediction}[c]) during episode two, and sampled a binary decision (action 'a' or action 'b') according to the model's bias.
In the same way as in the main behavioral experiment, each agent repeated the experiment four times and we estimated the empirical action-selection bias as  the mean of the (simulated) behavioral data over all repetitions of all agents.
This mean value depends on the actual realizations of the random variables and its uncertainty is higher when fewer samples are available.
We therefore repeated the simulation of $N=10$ agents $1000$ times and plot the distribution of the empirical means in Fig.~\ref{fig:PSA_ModelPrediction}[d].
The same procedure was repeated for $N=20$ agents, showing a smaller standard deviation.
The simulations showed a relatively large (simulated) effect size at states $D1$ and $D2$.
Furthermore, 
as expected, the action bias decays as a function of the delay between the action and the final reward in episode one.
We then compared the $Q-\lambda$ model with a member of the class of \textit{RL without eligibility trace}. 
When the parameter $\lambda$, which controls the decay of the eligibility trace, is set to $0$,  $Q-\lambda$ turns into \textit{$Q-$Learning without eligibility trace} and we can use it  to compare the two classes of RL without changing other parameters. 
Thus, we repeated the simulation for this case ($\lambda = 0$,  $N=20$) which shows the model predictions under our null hypothesis. Figure~\ref{fig:PSA_ModelPrediction}[d] shows the qualitative difference between the two classes of RL.

\subsection{Regression Analysis}

\begin{figure*}
	\centering
	\includegraphics[width=\textwidth]{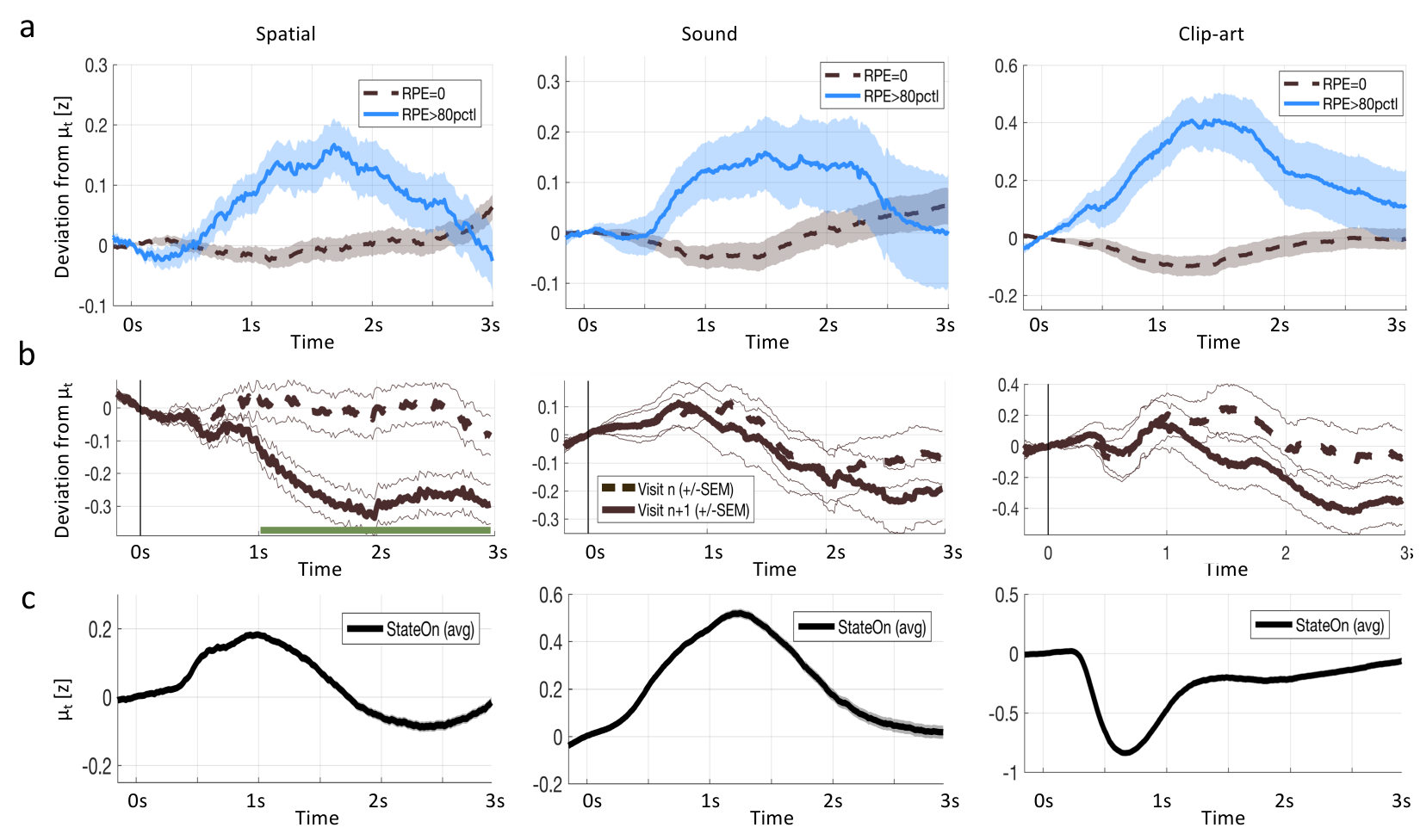}
	\caption{\textbf{Reward prediction error (RPE) at non-goal states modulates pupil dilation.}
	  Pupil traces (in units of standard deviation) from all states except G
          were aligned at state onset ($t=0ms$) and the mean pupil response $\mu_t$ was subtracted (see Methods).
          	\textbf{[a]}
		The deviation from the mean is shown for states where the model predicts $RPE = 0$ (black, dashed) and for states where the model predicts $RPE \geq 80^{th}$ percentile (solid, blue).
		Shaded areas: $\pm$ SEM.
		Thus the pupil dilation reflects the RPE
                predicted by a reinforcement learning model that spreads value information to nonrewarded states via eligibility traces.
\textbf{[b]}                
          To qualitatively distinguish pupil correlations with RPE from correlations with state values $V(s)$, we started from the following observation:
 the model predicts that RPE decreases over the course of learning (due to convergence), while the state values $V(s)$ increase (due to spread of value information).
	We wanted to observe this qualitative difference in the pupil dilations of subsequent visits of the \textit{same} state. 
	We selected pairs of visits $n$ and $n+1$ for which the RPE decreased while $V(s)$ increased and extracted the pupil measurements of the two visits (again, mean $\mu_t$ is subtracted). 
	The dashed, black curves show the average pupil trace during the $n^{th}$ visit of a state.
	The solid black curves correspond to the next visit ($n+1$) of the same state.
	In the spatial condition, the two curves significantly ($p<FDR_{\alpha}=0.05$) separate at $t>1s$ (indicated by the green line).
	All three conditions show the same trend (with strong significance in the spatial condition), compatible with a positive correlation of pupil response with RPE, but not with state value $V(s)$.
	\textbf{[c]} The mean pupil dilation $\mu_t$ is different in each condition, whereas the learning related deviations from the mean (in [a] and [b]) have similar shapes.
		}
	\label{fig:RPE_Regression_Raw}
\end{figure*}

The reward prediction error (RPE, Eq. \ref{eq:RewardPredictionError}) used for a comparison with pupil data
was obtained by applying the algorithm  \textit{Q-$\lambda$} with the optimal (maximum likelihood) parameters.
We chose \textit{Q-$\lambda$} for regression because, first, it explained the behavior best across the three conditions and, second, it evaluates the outcome of an action at the onset of the next state (rather than at the selection of the next action as in \textit{SARSA-$\lambda$}), which enabled us to compare the model with the pupil traces triggered at the onset of the next state.

In a first, qualitative, analysis, we split data of all state transitions of all Participants into two groups:
all the state transitions where the model predicts an RPE of zero;
and the twenty percent of state transitions where the model predicts the largest  RPE (Fig.~\ref{fig:RPE_Regression_Raw}[a]). 
We found that the pupil responses looked very different in the two groups, across all three modalities.

\begin{figure*}
\centering
	\includegraphics[width=\textwidth]{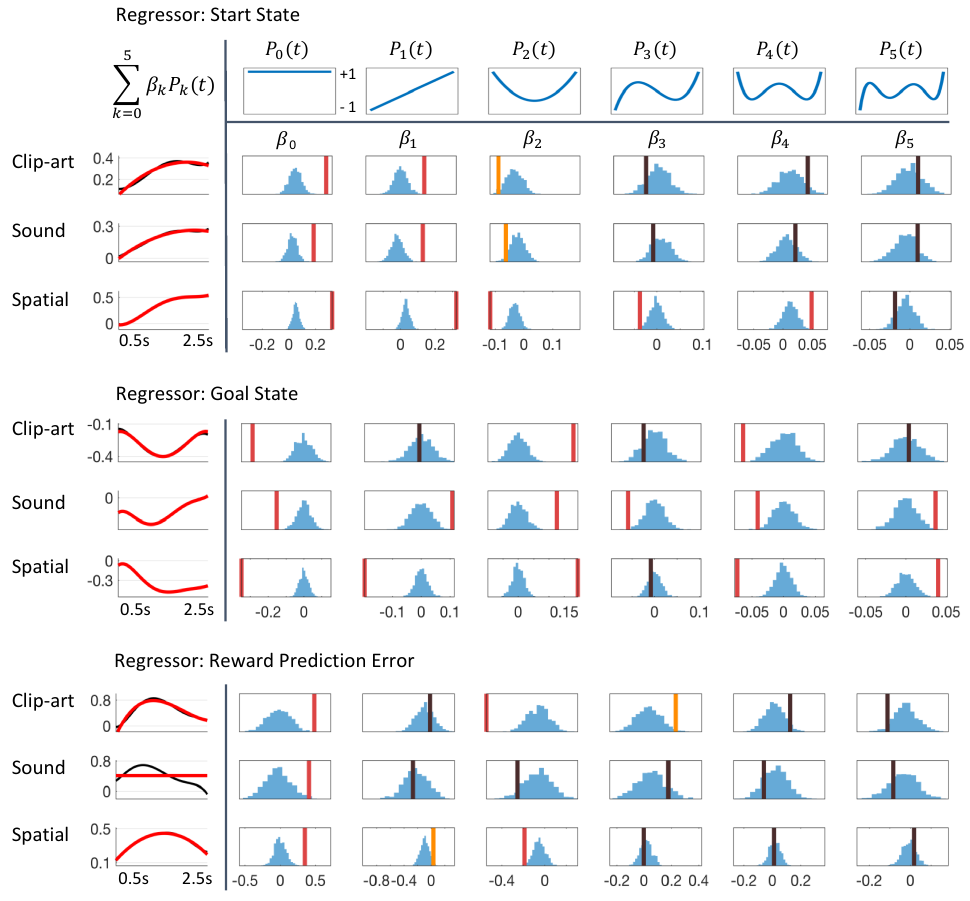}
	\caption{\textbf{Detailed results of regression analysis and permutation tests.}
	The regressors are \textit{top}: Start state event, \textit{middle}: Goal state event and \textit{bottom}: Reward Prediction Error.
	We extracted the time course of the pupil dilation in [$500ms$, $2500ms$] after state onset for each of the  conditions, \textit{clip-art}, \textit{sound} and \textit{spatial}, using Legendre polynomials  $P_k(t)$ of orders k=0 to k=5 (top row) as basis functions.
	The extracted weights $\beta_k$ (cf. Eq. \ref{eq:LegendreRegression}) are shown in each column below the corresponding Legendre polynomial as vertical bars with color indicating the level of significance
	(red, statistically significant at p<0.05/6 (Bonferroni); orange, p<0.05; black, not significant).
	Blue histograms summarize shuffled samples obtained by 1000 permutations.
	Black curves in the leftmost column show the fits with all 6 Legendre Polynomials,
	while the red curve is obtained by summing only over the few Legendre Polynomials with significant $\beta$.
	Note the similarity of the pupil responses across conditions.}
	\label{fig:RegressionAndPermTest}
\end{figure*}

In a second, rigorous, statistical analysis, we  tested whether pupil responses were correlated  with the RPE across all RPE values, not just those in the two groups with zero and very high RPE.
In our experiment, only state 'G' was rewarded; at non goal states, the RPE depended solely on learned $Q$-values ($r_{n+1} = 0$ in Eq. \ref{eq:RewardPredictionError}).
Note that at the first state of each episode the RPE is not defined.
We distinguished these three cases in the regression analysis by defining two events "Start" and "Goal", as well as a parametric modulation by the reward prediction error at intermediate states.
From Figure \ref{fig:PupilDiameter_GrandAvg} we expected significant modulations in the time window $t \in [500ms, 2500ms]$ after stimulus onset.
We mapped $t$ to $t' = (t-1500ms)/1000ms$ and used orthogonal Legendre polynomials $P_k(t')$ up to order $k=5$ (Fig.~\ref{fig:RegressionAndPermTest}) as basis functions on the interval $-1 < t' < 1$.
We use the indices $p$ for participant and $n$ for the $n^{th}$ state-on event. With a noise term $\epsilon$ and $\mu_t$ for the overall mean pupil dilation at $t$, the regression model for the pupil measurements $y$ is
\begin{equation}
y_{p, n+1, t} = \mu_t + \sum_{k=0}^{5} RPE_{p}(n \rightarrow n+1) \times P_{k}(t') \times \beta_k  \; + \epsilon_{p,n+1,t}  \, ,
	\label{eq:LegendreRegression}
\end{equation}
where the participant-independent parameters $\beta_k$ were fitted to the experimental data (one independent analysis for each experimental condition).
The models for "start state" and "goal state" are analogous and obtained by replacing the real valued $RPE_{p, n} $ by a 0/1 indicator for the respective events.
By this design we obtained three uncorrelated regressors with six parameters each.

Using the regression analysis sketched here, we quantified the qualitative observations
suggested by (Fig.~\ref{fig:RPE_Regression_Raw}) 
and found a significant parametric modulation of the pupil dilation by reward prediction errors at non-goal states (Fig.~\ref{fig:RegressionAndPermTest}).
The extracted modulation profile reached a maximum at around $1-1.5s$ ( 1300 ms in the \textit{clip-art},  1100 ms in the \textit{sound} and 1400 ms in the \textit{spatial} condition), with a strong mean effect size ($\beta_0$ in Fig.~\ref{fig:RegressionAndPermTest}) of $0.48$ ($p<0.001$),   $0.41$ ($p =0.008$) and $0.35$ ($p<0.001$), respectively.

We interpret the pupil traces at the start and the end of each episode (Fig.~\ref{fig:RegressionAndPermTest})
as markers for additional cognitive processes beyond reinforcement learning
which could include correlations 
with cognitive load \citep{Kahneman66, Beatty82}, recognition memory \citep{Otero11},  attentional effort \citep{Alnaes14}, exploration \citep{Jepma11}, and encoding of memories \citep{Kucewicz18}.

\clearpage

\bibliographystyle{unsrt}

\bibliography{ETBibliography}

\begin{thebibliography}{10}

\bibitem{Sutton18}
Richard~S. Sutton and Andrew~G. Barto.
\newblock {\em Reinforcement Learning: An Introduction}.
\newblock MIT Press, Cambridge, MA, (in progress) second edition, 2018.

\bibitem{Pessiglione06}
Mathias Pessiglione, Ben Seymour, Guillaume Flandin, Raymond~J. Dolan, and
  Chris~D. Frith.
\newblock {Dopamine-dependent prediction errors underpin reward-seeking
  behaviour in humans}.
\newblock {\em Nature}, 442(7106):1042--1045, 2006.

\bibitem{Glaescher10}
Jan Gl{\"a}scher, Nathaniel Daw, Peter Dayan, and John~P. O'Doherty.
\newblock States versus rewards: Dissociable neural prediction error signals
  underlying model-based and model-free reinforcement learning.
\newblock {\em Neuron}, 66(4):585--595, 2010.

\bibitem{Daw11}
Nathaniel~D. Daw, Samuel~J. Gershman, Ben Seymour, Peter Dayan, and Raymond~J.
  Dolan.
\newblock {Model-based influences on humans' choices and striatal prediction
  errors}.
\newblock {\em Neuron}, 69(6):1204--1215, 2011.

\bibitem{Niv12}
Yael Niv, J.~A. Edlund, Peter Dayan, and John~P. O'Doherty.
\newblock {Neural Prediction Errors Reveal a Risk-Sensitive
  Reinforcement-Learning Process in the Human Brain}.
\newblock {\em Journal of Neuroscience}, 32(2):551--562, 2012.

\bibitem{ODoherty17}
John~P. O'Doherty, Jeffrey Cockburn, and Wolfgang~M Pauli.
\newblock {Learning, Reward, and Decision Making}.
\newblock {\em Annu. Rev. Psychol.}, 68:73--100, 2017.

\bibitem{Tartaglia17}
Elisa~M. Tartaglia, Aaron~M. Clarke, and Michael~H. Herzog.
\newblock {What to choose next? A paradigm for testing human sequential
  decision making}.
\newblock {\em Frontiers in Psychology}, 8:1--11, 2017.

\bibitem{Sutton88a}
Richard~S. Sutton.
\newblock Learning to predict by the methods of temporal differences.
\newblock {\em Machine Learning}, 3(1):9--44, 1988.

\bibitem{Watkins89}
C.J.C.H. Watkins.
\newblock {\em Learning from delayed rewards}.
\newblock PhD-thesis, Cambridge University, 1989.

\bibitem{Williams92}
R.J. Williams.
\newblock {Simple Statistical Gradient-Following Algorithms for Connectionist
  Reinforcement Learning}.
\newblock {\em Reinforcement Learning}, 8:229--256, 1992.

\bibitem{Peng96}
Jing Peng and Ronald~J Williams.
\newblock {Incremental Multi-Step Q-Learning}.
\newblock {\em Machine Learning}, 22:283--290, 1996.

\bibitem{Singh96}
Satinder Singh and Richard~S. Sutton.
\newblock {Reinforcement Learning with replacing elibibility traces}.
\newblock {\em Machine Learning}, 22:123--158, 1996.

\bibitem{Mnih16}
Volodymyr Mnih, Adria~Puigdomenech Badia, Mehdi Mirza, Alex Graves, Timothy
  Lillicrap, Tim Harley, David Silver, and Koray Kavukcuoglu.
\newblock Asynchronous methods for deep reinforcement learning.
\newblock In {\em Proceedings of The 33rd International Conference on Machine
  Learning}, 2016.

\bibitem{Moore93}
Andrew~W. Moore and Christopher~G. Atkeson.
\newblock Prioritized sweeping: Reinforcement learning with less data and less
  time.
\newblock {\em Machine Learning}, 13(1):103--130, 1993.

\bibitem{Blundell16}
C.~{Blundell}, B.~{Uria}, A.~{Pritzel}, Y.~{Li}, A.~{Ruderman}, J.~Z {Leibo},
  J.~{Rae}, D.~{Wierstra}, and D.~{Hassabis}.
\newblock {Model-Free Episodic Control}.
\newblock {\em ArXiv e-prints}, 2016.

\bibitem{Yagishita14}
S.~Yagishita, A.~Hayashi-Takagi, G.~C.~R. Ellis-Davies, H.~Urakubo, S.~Ishii,
  and H.~Kasai.
\newblock A critical time window for dopamine actions on the structural
  plasticity of dendritic spines.
\newblock {\em Science}, 345(6204):1616--1620, 2014.

\bibitem{He15}
Kaiwen He, Marco Huertas, Su~Z. Hong, XiaoXiu Tie, Johannes~W. Hell, Harel
  Shouval, and Alfredo Kirkwood.
\newblock Distinct eligibility traces for ltp and ltd in cortical synapses.
\newblock {\em Neuron}, 88(3):528--538, 2015.

\bibitem{Bittner17}
Katie~C Bittner, Aaron~D. Milstein, Christine Grienberger, Sandro Romani, and
  Jeffrey~C. Magee.
\newblock {Behavioral time scale synaptic plasticity underlies CA1 place
  fields}.
\newblock {\em Science}, 357(6355):1033--1036, 2017.

\bibitem{Fisher17}
Simon~D. Fisher, Paul~B. Robertson, Melony~J. Black, Peter Redgrave, Mark~A.
  Sagar, Wickliffe~C. Abraham, and John~N.J. Reynolds.
\newblock {Reinforcement determines the timing dependence of corticostriatal
  synaptic plasticity in vivo}.
\newblock {\em Nature Communications}, 8(1), 2017.

\bibitem{Gerstner18}
Wulfram Gerstner, Marco Lehmann, Vasiliki Liakoni, Dane Corneil, and Johanni
  Brea.
\newblock {Eligibility Traces and Plasticity on Behavioral Time Scales:
  Experimental Support of NeoHebbian Three-Factor Learning Rules}.
\newblock {\em Frontiers in Neural Circuits}, 12:53, 2018.

\bibitem{Walsh11}
Matthew~M. Walsh and John~R. Anderson.
\newblock {Learning from delayed feedback: Neural responses in temporal credit
  assignment}.
\newblock {\em Cognitive, Affective and Behavioral Neuroscience},
  11(2):131--143, 2011.

\bibitem{ODoherty03}
J~O'Doherty, P~Dayan, K~Friston, H~Critchley, and R~Dolan.
\newblock Temporal difference learning model accounts for responses in human
  ventral striatum and orbitofrontal cortex during pavlovian appetitive
  learning.
\newblock {\em Neuron}, 38:329--337, 2003.

\bibitem{Jepma11}
Marieke Jepma and Sander Nieuwenhuis.
\newblock {Pupil diameter predicts changes in the exploration-exploitation
  trade-off: evidence for the adaptive gain theory}.
\newblock {\em Journal of cognitive neuroscience}, 23:1587--1596, 2011.

\bibitem{Otero11}
Samantha~C. Otero, Brendan~S. Weekes, and Samuel~B. Hutton.
\newblock {Pupil size changes during recognition memory}.
\newblock {\em Psychophysiology}, 48(10):1346--1353, 2011.

\bibitem{Preuschoff11}
Kerstin Preuschoff, Bernard~Marius {'t Hart}, and Wolfgang Einh{\"{a}}user.
\newblock {Pupil dilation signals surprise: Evidence for noradrenaline's role
  in decision making}.
\newblock {\em Frontiers in Neuroscience}, 5:1--12, 2011.

\bibitem{Bogacz07b}
Rafal Bogacz, Samuel~M. McClure, Jian Li, Jonathan~D Cohen, and P.~Read
  Montague.
\newblock {Short-term memory traces for action bias in human reinforcement
  learning}.
\newblock {\em Brain Research}, 1153(1):111--121, 2007.

\bibitem{Walsh12}
Matthew~M. Walsh and John~R. Anderson.
\newblock {Learning from experience: Event-related potential correlates of
  reward processing, neural adaptation, and behavioral choice}.
\newblock {\em Neuroscience and Biobehavioral Reviews}, 36(8):1870--1884, 2012.

\bibitem{Weinberg12}
Anna Weinberg, Christian~C. Luhmann, Jennifer~N. Bress, and Greg Hajcak.
\newblock {Better late than never? The effect of feedback delay on ERP indices
  of reward processing}.
\newblock {\em Cognitive, Affective and Behavioral Neuroscience},
  12(4):671--677, 2012.

\bibitem{Rescorla72}
R.A. Rescorla and A.R. Wagner.
\newblock A theory of {P}avlovian conditioning: variations in the effectiveness
  of reinforcement and nonreinforcement.
\newblock In {\em Classical Conditioning \protect{II}: current research and
  theory}. Appleton Century Crofts, 1972.

\bibitem{Yates66}
Frances~A. Yates.
\newblock {\em {Art of Memory}}.
\newblock Routledge and Kegan Paul, 1966.

\bibitem{Benjamini95}
Y.~Benjamini and Y.~Hochberg.
\newblock {Controlling the False Discovery Rate : A Practical and Powerful
  Approach to Multiple Testing}.
\newblock {\em J. R. Statist. Soc.}, 57(1):289--300, 1995.

\bibitem{Glimcher13}
Paul~W. Glimcher and Ernst Fehr, editors.
\newblock {\em {Neuroeconomics: Decision Making and the Brain: Second
  Edition}}.
\newblock Elsevier Inc., 2 edition, 2013.

\bibitem{Akaike74}
Hirotugu Akaike.
\newblock {A New Look at the Statistical Model Identification}.
\newblock {\em IEEE Trans. Autom. Control AC-19}, 19:716--723, 1974.

\bibitem{Burnham04}
Kenneth~P. Burnham and David~R. Anderson.
\newblock {Multimodel inference: Understanding AIC and BIC in model selection}.
\newblock {\em Sociological Methods and Research}, 33(2):261--304, 2004.

\bibitem{Standing73}
Lionel Standing.
\newblock {Learning 10000 Pictures}.
\newblock {\em The Quarterly journal of experimental psychology}, 25:207--222,
  1973.

\bibitem{Brady08}
T.~F. Brady, T.~Konkle, G.~A. Alvarez, and A.~Oliva.
\newblock {Visual long-term memory has a massive storage capacity for object
  details}.
\newblock {\em Proceedings of the National Academy of Sciences},
  105(38):14325--14329, 2008.

\bibitem{Duncan16}
Katherine~D Duncan and Daphna Shohamy.
\newblock {Memory States Influence Value-Based Decisions}.
\newblock {\em Journal of Experimental Psychology: General},
  145(11):1420--1426, 2016.

\bibitem{Greve17}
Andrea Greve, Elisa Cooper, Alexander Kaula, Michael~C. Anderson, and Richard
  Henson.
\newblock {Does prediction error drive one-shot declarative learning?}
\newblock {\em Journal of memory and language}, 94:149--165, 2017.

\bibitem{Rouhani18}
Nina Rouhani, Kenneth~A. Norman, and Yael Niv.
\newblock {Dissociable effects of surprising rewards on learning and memory.}
\newblock {\em Journal of Experimental Psychology: Learning, Memory, and
  Cognition}, 44(9):1430--1443, 2018.

\bibitem{Pan05}
Wei-xing Pan, Robert Schmidt, Jeffery~R Wickens, and Brian~I Hyland.
\newblock {Dopamine cells respond to predicted events during classical
  conditioning: evidence for eligibility traces in the reward-learning
  network}.
\newblock {\em Journal of Neuroscience}, 25(26):6235--6242, 2005.

\bibitem{Gureckis09}
Todd~M. Gureckis and Bradley~C. Love.
\newblock {Short-term gains, long-term pains: How cues about state aid learning
  in dynamic environments}.
\newblock {\em Cognition}, 113(3):293--313, 2009.

\bibitem{Crow68}
T.~Crow.
\newblock Cortical synapses and reinforcement: a hypothesis.
\newblock {\em Nature}, 219:736--737, 1968.

\bibitem{Fremaux16}
Nicolas Fr\'{e}maux and Wulfram Gerstner.
\newblock Neuromodulated spike-timing-dependent plasticity, and theory of
  three-factor learning rules.
\newblock {\em Frontiers in Neural Circuits}, 9, 2016.

\bibitem{Seijen13}
Harm~Van Seijen and Rich Sutton.
\newblock Planning by prioritized sweeping with small backups.
\newblock In {\em Proceedings of the 30th International Conference on Machine
  Learning}, 2013.

\bibitem{Brea17a}
J.~{Brea}.
\newblock {Is prioritized sweeping the better episodic control?}
\newblock {\em ArXiv e-prints}, 2017.

\bibitem{Izhikevich07}
Eugene~M Izhikevich.
\newblock {\em Dynamical systems in neuroscience : the geometry of excitability
  and bursting}.
\newblock MIT Press, 2007.

\bibitem{Brzosko17}
Zuzanna Brzosko, Sara Zannone, Wolfram Schultz, Claudia Clopath, and Ole
  Paulsen.
\newblock {Sequential neuromodulation of hebbian plasticity offers mechanism
  for effective reward-based navigation}.
\newblock {\em eLife}, 6:27756, 2017.

\bibitem{Schultz15}
Wolfram Schultz.
\newblock {Neuronal Reward and Decision Signals: From Theories to Data}.
\newblock {\em Physiological Reviews}, 95(3):853--951, 2015.

\bibitem{Kucewicz18}
Michal~T. Kucewicz, Jaromir Dolezal, Vaclav Kremen, Brent~M. Berry, Laura~R.
  Miller, Abigail~L. Magee, Vratislav Fabian, and Gregory~A. Worrell.
\newblock {Pupil size reflects successful encoding and recall of memory in
  humans}.
\newblock {\em Scientific Reports}, 8(1):4949, 2018.

\bibitem{Joshi16}
Siddhartha Joshi, Yin Li, Rishi~M. Kalwani, and Joshua~I. Gold.
\newblock {Relationships between Pupil Diameter and Neuronal Activity in the
  Locus Coeruleus, Colliculi, and Cingulate Cortex}.
\newblock {\em Neuron}, 89(1):221--234, 2016.

\bibitem{Berke18}
Joshua~D. Berke.
\newblock {What does dopamine mean?}
\newblock {\em Nature Neuroscience}, 21(6):787--793, 2018.

\bibitem{Nieuwenhuis11}
Sander Nieuwenhuis, Eco {De Geus}, and Gary Aston-jones.
\newblock {The anatomical and functional relationship between the P3 and
  autonomic components of the orienting response}.
\newblock {\em Psychophysiology}, 48:162--175, 2011.

\bibitem{Brainard97}
{Brainard D. H.}
\newblock {The Psychophysics Toolbox}.
\newblock {\em Spatial Vision}, 10:433--436, 1997.

\bibitem{Mathot17}
Sebastiaan Math{\^{o}}t, Jasper Fabius, Elle {Van Heusden}, and Stefan {Van der
  Stigchel}.
\newblock {Safe and sensible baseline correction of pupil-size data}.
\newblock {\em PeerJ PrePrints}, pages 1--25, 2017.

\bibitem{Hastings70}
W.~K. Hastings.
\newblock {Monte Carlo simulation methods using Markov Chains and their
  applications}.
\newblock {\em Biometrika}, 57:97--109, 1970.

\bibitem{Kahneman66}
D~Kahneman and J~Beatty.
\newblock {Pupil diameter and load on memory.}
\newblock {\em Science (New York, N.Y.)}, 154(3756):1583--5, 1966.

\bibitem{Beatty82}
Jackson Beatty.
\newblock {Task-evoked pupillary responses, processing load, and the structure
  of processing resources}.
\newblock {\em Psychological Bulletin}, 91(2):276--292, 1982.

\bibitem{Alnaes14}
D.~Alnaes, M.~H. Sneve, T.~Espeseth, T.~Endestad, S.~H.~P. van~de Pavert, and
  B.~Laeng.
\newblock {Pupil size signals mental effort deployed during multiple object
  tracking and predicts brain activity in the dorsal attention network and the
  locus coeruleus}.
\newblock {\em Journal of Vision}, 14(4), 2014.

\end{thebibliography}

\end{document}